\documentclass[reprint,amsmath,amssymb,aps,prl,floatfix]{revtex4-1}

\usepackage{graphicx}
\usepackage{dcolumn}
\usepackage{bbold}
\usepackage{color}
\usepackage{booktabs}
\usepackage[dvipsnames]{xcolor}
\usepackage{soul}
\usepackage{tikz}
\usepackage{hyperref}

\newcommand{\SU}{\mathrm{SU}}

\newcommand{\dg}{\dagger}

\newcommand{\one}{\mathbb{1}}

\newcommand{\Tr}{\mathrm{Tr}}
\newcommand{\mg}[2]{\vcenter{\hbox{\includegraphics[scale=#1]{#2}}}}

\newcommand{\LGCNN}{{L-CNN}}
\newcommand{\LConv}{L-Conv}
\newcommand{\LBL}{{L-Bilin}}
\newcommand{\LAct}{{L-Act}}
\newcommand{\LExp}{{L-Exp}}
\newcommand{\Plaq}{Plaq}
\newcommand{\Poly}{Poly}
\newcommand{\Trace}{Trace}
\newcommand{\LGE}{LGE}

\begin{document}

\title{Lattice gauge equivariant convolutional neural networks}

\author{Matteo Favoni}
\email{favoni@hep.itp.tuwien.ac.at}
\author{Andreas Ipp}
\email{ipp@hep.itp.tuwien.ac.at}
\author{David I.~M\"uller}
\email[Corresponding author: ]{dmueller@hep.itp.tuwien.ac.at}
\author{Daniel Schuh}
\email{schuh@hep.itp.tuwien.ac.at}
\affiliation{Institute for Theoretical Physics, TU Wien, Austria}

\date{\today}

\begin{abstract}
We propose Lattice gauge equivariant Convolutional Neural Networks (\LGCNN{}s) for generic machine learning applications on lattice gauge theoretical problems. At the heart of this network structure is a novel convolutional layer that preserves gauge equivariance while forming arbitrarily shaped Wilson loops in successive bilinear layers. Together with topological information, for example from Polyakov loops, such a network can in principle approximate any gauge covariant function on the lattice. We demonstrate that \LGCNN{}s can learn and generalize gauge invariant quantities that traditional convolutional neural networks are incapable of finding.
\end{abstract}

\maketitle

Gauge field theories are an important cornerstone of modern physics and encompass the fundamental forces of nature, including electromagnetism and nuclear forces. The physical information is captured in Wilson loops~\cite{Wilson:1974sk}, or holonomies, which describe how a quantity is parallel transported along a given closed path. Local gauge transformations can modify the fundamental fields independently at each space-time point but leave any traced Wilson loop invariant. On the lattice, gauge invariant observables are typically formulated in terms of traced Wilson loops of different shapes. The most basic example is the Wilson action which is formulated entirely in terms of \mbox{$1 \times 1$} loops, so-called plaquettes. The Wilson action can be systematically improved by including terms involving larger loops~\cite{Niedermayer:1996eb,Iwasaki:1985we,Moore:1996wn,Lagae:1998pe,Ipp:2018hai}. Planar rectangular loops are used for characterizing confinement. Most famously, the potential of a static quark pair can be computed from the expectation value of a Wilson loop with large extent in the temporal direction~\cite{Bali:2000gf}. Improved approximations to the energy momentum tensor or the topological charge density can involve also non-planar loops of growing size~\cite{Caracciolo:1989pt,Alexandrou:2017hqw,BilsonThompson:2002jk}. As the number of possible loops on a lattice grows exponentially with its path length, a systematic treatment of higher order contributions can become increasingly challenging.

Artificial neural networks provide a way to automatically extract relevant information from large amounts of data. They have become increasingly popular in many Abelian lattice applications, such as for $\phi^4$ scalar field, Ising, XY, Potts or Yukawa models, where they can recognize classical~\cite{Zhou:2018ill} and topological~\cite{Wang:2020hji} phase transitions from field configurations, determine local and non-local features~\cite{Grimmer:2019ahl,Bachtis:2020ajb} or infer action parameters~\cite{Bluecher:2020kxq}. Neural networks can improve the efficiency of sampling techniques~\cite{Urban:2018tqv}, extract optimal renormalization group transformations~\cite{Hu:2019nea}, or reconstruct spectral functions from Green's functions~\cite{Kades:2019wtd}. By the universal approximation theorem, these networks can, in principle, learn any function~\cite{Cybenko:1989a,Lu:2017a,Zhou:2018a}. In order to avoid merely memorizing training samples, imposing additional restrictions on these networks can improve their generalization capabilities~\cite{kawaguchi:2020a}. Global translational equivariance induces convolutions~\cite{Jaehne:1997} which form the basis of convolutional neural networks (CNNs). Additional global symmetry groups, such as global rotations, can be incorporated using Group equivariant CNNs (G-CNNs)~\cite{Cohen:2016a,Kondor:2018a,Cheng:2019xrt,Esteves:2020a,Rath:2020a,Gerken:2021sla}. This approach can be extended to local gauge symmetries. Even though gauge-invariant observables can be learned to some extent by non-equivariant networks~\cite{Boyda:2020nfh}, recently there has been a lot of interest in incorporating gauge symmetries directly into the network structure. For discrete ones, equivariant network structures have been implemented for the icosahedral group~\cite{Cohen:2019xsh} or for the $\mathbb{Z}_2$ gauge group~\cite{Luo:2020stn}; for continuous ones, a much larger symmetry space is available~\cite{Finzi:2020a}. A recent seminal work demonstrated that incorporating $\mathrm{U}(1)$ or $\SU(N_c)$ gauge symmetries into a neural network can render flow-based sampling orders of magnitude faster than traditional approaches~\cite{Kanwar:2020xzo,Boyda:2020hsi}. This impressive result was obtained using parametrized invertible coupling layers that essentially depend on parallel-transported plaquettes. Up to now, machine learning applications that require larger Wilson loops have relied on manually picking a set of relevant Wilson loops~\cite{Shanahan:2018vcv} or on simplifications due to the choice of a discrete Abelian gauge group~\cite{Zhang:2019a}. A comprehensive treatment for continuous non-Abelian gauge groups has been missing so far, and there is an obvious desire to systematically generate all Wilson loops from simple local operations.

In this Letter, we introduce Lattice Gauge Equivariant (LGE) CNNs (abbreviated \LGCNN{}s), which we intend as a gauge equivariant replacement for traditional CNNs in machine learning problems for lattice gauge theory. We specify a basic set of network layers that preserve gauge symmetry exactly while allowing for universal expressivity for physically distinct field configurations. Gauge equivariant layers can be stacked arbitrarily to form gauge equivariant networks. In particular, we provide a new convolutional operation which, in combination with a gauge equivariant bilinear layer, can grow arbitrarily shaped Wilson loops from local operations. We show that the full set of all contractible Wilson loops can be constructed in this way. Together with topological information from non-contractible loops, in principle, the full gauge connection can be reconstructed~\cite{Giles:1981ej,Loll:1992fk}. Trace layers produce gauge invariant output that can be linked to physical observables. Using simple regression tasks for Wilson loops of different sizes and shapes in pure SU($2$) gauge theory, we demonstrate that \LGCNN{}s outperform conventional CNNs by far, especially with growing loop size.

\begin{figure*}
    \centering
    \includegraphics[width=0.75\textwidth]{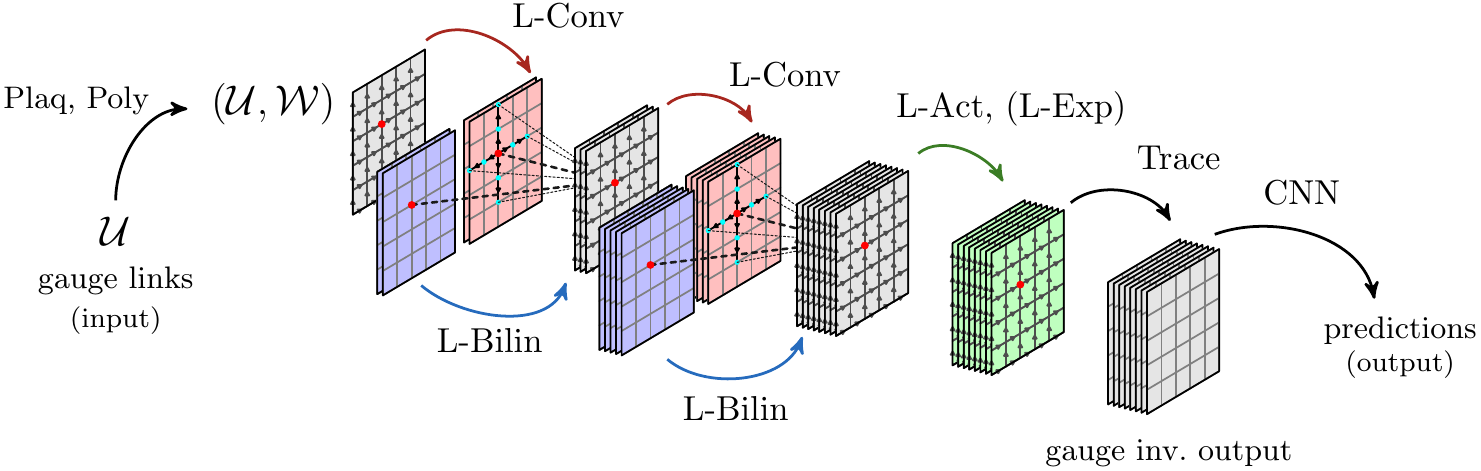}
    \caption{
        A possible realization of an \LGCNN{}. Lattice data in the form of $\mathcal{U}$ links are first preprocessed by \Plaq{} and \Poly{} in order to generate elementary locally transforming $\mathcal{W}$ objects. 
        An \LConv{} is used to parallel transport nearby $\mathcal{W}$ objects (green dots) along the coordinate axes to a particular lattice site (red dot).
        An \LBL{} combines two layers by forming products of locally transforming objects which are stored in an increasing number of channels (indicated by stacked lattices).
        The second input layer (blue) for this operation can be a duplicate of the original layer (red).
        An additional \LAct{} (\LExp{}) can modify $\mathcal W$ ($\mathcal U$) in a gauge equivariant way (green layer).
        A \Trace{} layer generates gauge invariant output that can be further processed by a traditional CNN.
        The example depicts a 1+1D lattice but applies to higher dimensions as well.
        The basic layers presented can be combined to form other deeper network architectures.}
    \label{fig:layers}
\end{figure*}

Lattice gauge theory is a discretization of Yang-Mills theory~\cite{Wilson:1974sk,Gattringer:2010abc, Smit:2002aaa}. We consider a system at finite temperature with gauge group $\SU(N_c)$ in $D+1$ dimensions on a lattice $\Lambda$ of size $N_t \cdot N_s^D$ with $N_t$ ($N_s$) cells along the imaginary time (spatial) direction(s) with periodic boundary conditions. The link variables $U_{x,\mu}$ specify the parallel transport from a lattice site $x$ to its neighbor $x + \mu \equiv x + a \hat{e}^\mu$  with lattice spacing $a$. Gauge links transform according to
\begin{align} \label{eq:link_tr}
    T_\Omega U_{x,\mu} = \Omega_x U_{x,\mu} \Omega^\dg_{x+\mu},
\end{align}
where the group elements $\Omega_x$ are unitary and have unit determinant. The Yang-Mills action can be approximated by the Wilson action~\cite{Wilson:1974sk}
\begin{align} \label{eq:wilson_action}
    S_W[U] = \frac{2}{g^2} \sum_{x \in \Lambda} \sum_{\mu < \nu} \mathrm{Re} \, \Tr \left[ \one - U_{x,\mu\nu} \right]
\end{align}
with the plaquette variables
\begin{align}
    U_{x,\mu\nu} = U_{x,\mu} U_{x+\mu, \nu} U^\dg_{x+\nu, \mu} U^\dg_{x,\nu} = \mg{0.7}{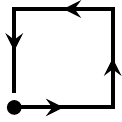} \label{eq:plaquette}
\end{align}
which are \mbox{$1 \times 1$} (untraced) Wilson loops on the lattice. Unless specified otherwise, we assume Wilson loops to be untraced, i.e.~matrix valued. The plaquette variables transform locally at $x$ as $T_\Omega U_{x,\mu\nu} = \Omega_x U_{x,\mu\nu} \Omega^\dg_x.$

\LGCNN{}s can express a large class of possible gauge equivariant functions in the lattice gauge theory framework. As customary in feed-forward CNNs, we split \LGCNN{}s into more elementary ``layers'', see Fig.~\ref{fig:layers}. As input data for a layer we use a tuple $(\mathcal{U}, \mathcal{W})$ consisting of non-locally transforming gauge link variables $\mathcal{U}$ and locally transforming variables $\mathcal{W}$. The first part of the tuple is the set of variables $\mathcal{U} = \{ U_{x,\mu} \}$, which transform according to Eq.~\eqref{eq:link_tr}. For concreteness, we choose the defining (or fundamental) representation of $\SU(N_c)$ such that we can treat link variables as complex special unitary \mbox{$N_c \times N_c$} matrices.  Its second part is a set of variables $\mathcal{W} = \{ W_{x,i} \}$ with $W_{x,i} \in \mathbb C^{N_c \times N_c}$ and index $1 \leq i \leq N_\mathrm{ch}$, which we interpret as ``channels''. We require these additional input variables to transform locally at $x$:
\begin{align}
    T_\Omega W_{x,i} = \Omega_x W_{x,i} \Omega^\dg_x.
\end{align}

A function~$f$ that performs some mathematical operation on $( \mathcal{U}, \mathcal{W} )$ is called gauge equivariant (or gauge covariant) if  $f(T_\Omega \mathcal{U}, T_\Omega \mathcal{W}) = T'_\Omega f(\mathcal{U}, \mathcal{W})$, where $T'_\Omega f$ denotes the gauge transformed expression of the function~$f$. Additionally, a function~$f$ is gauge invariant if $f(T_\Omega \mathcal{U}, T_\Omega \mathcal{W}) = f(\mathcal{U}, \mathcal{W})$. All possible functions that can be expressed as \LGCNN{}s should either be equivariant or invariant.

\textbf{\LGE{} convolutions (\LConv)} perform a parallel transport of $\mathcal W$ objects at neighboring sites to the current location. They can be written as
\begin{align}
    W_{x, i} \rightarrow \sum_{j, \mu, k} \omega_{i, j, \mu, k} U_{x, k\cdot \mu} W_{x + k\cdot \mu, j} U^\dagger_{x, k\cdot \mu},
\end{align}
where $\omega_{i,j,\mu,k} \in \mathbb C$ are the weights of the convolution with $1 \le i \le N_\mathrm{ch,out}$, $1 \le j \le N_\mathrm{ch,in}$, $0 \le \mu \le D$ and \mbox{$-K$}$ \le k \le K$, where $K$ is the kernel size. Unlike traditional convolutional layers, the gauge equivariant kernels connect to other lattice sites only along the coordinate axes. The reason is path dependence. In the continuum case a natural choice would be the shortest path (or geodesic) connecting $x$ and $y$, which is also used for gauge equivariant neural networks that are formulated on manifolds~\cite{Cohen:2019xsh}. However, in our lattice approach the shortest path is not unique, unless one restricts oneself to the coordinate axes. Possible variations of this layer are to include an additional bias term, or to restrict to even sparser dilated convolutions~\cite{Yu:2016jsd}.

\textbf{\LGE{} bilinear layers (\LBL)} combine two tuples $(\mathcal{U}, \mathcal{W})$ and $(\mathcal{U}, \mathcal{W}')$ to form products of locally transforming quantities as
\begin{align} \label{eq:gcbl}
    W_{x,i} \rightarrow \sum_{j,k} \alpha_{i,j,k} W_{x,j} W'_{x,k},
\end{align}
where $\alpha_{i,j,k} \in \mathbb C$ are parameters with $1 \le i \le N_\mathrm{out}$, $1 \le j \le N_\mathrm{in, 1}$ and $1 \le k \le N_\mathrm{in, 2}$. Since only locally transforming terms are multiplied in Eq.~\eqref{eq:gcbl}, gauge equivariance holds. For more flexibility, the bilinear operation can be further generalized by enlarging $\mathcal{W}$ and $\mathcal{W}'$ to also include the unit element $\one$ and all Hermitian conjugates of $\mathcal{W}$ and $\mathcal{W}'$. An \LBL{} can then also act as residual module~\cite{He:2016res} and includes a bias term.

\textbf{\LGE{} activation functions (\LAct)} can be applied at each lattice site via
\begin{align}
    W_{x,i} \rightarrow g_{x,i}(\mathcal{U}, \mathcal{W}) W_{x,i}\label{eq:activationfunction}
\end{align}
using any scalar-valued, gauge invariant function $g$. A gauge equivariant generalization of the commonly used rectified linear unit (ReLU)  could be realized by choosing $g_{x,i}(\mathcal{U}, \mathcal{W}) = \mathrm{ReLU}(\mathrm{Re}\, \Tr \left[ W_{x,i} \right])$ where $g$ only depends on local variables. In general, $g$ can depend on values of variables at any lattice site and, in principle, could also depend on trainable parameters.

\textbf{\LGE{} exponentiation layers (\LExp)} can be used to update the link variables through
\begin{align} \label{eq:exp_layer_U}
    U_{x,\mu} \rightarrow U'_{x,\mu} = \mathcal{E}_{x,\mu} U_{x,\mu},
\end{align}
where $\mathcal{E}_{x,\mu} \in \SU(N_c)$ is a group element which transforms locally $T_\Omega \mathcal{E}_{x,\mu} = \Omega_x \mathcal{E}_{x,\mu} \Omega^\dg_x$. By this update, the unitarity (${U'}^{\dg}_{x,\mu} U'_{x,\mu} = \one$) and determinant ($\det U'_{x,\mu} = 1$) constraints remain satisfied. A particular realization of $\mathcal{E}_{x,\mu}$ in terms of $\mathcal{W}$-variables is given by the exponential map
\begin{align} \label{eq:exp_layer}
    \mathcal{E}_{x,\mu}(\mathcal{W}) = \exp{ \left( i  \sum_i \beta_{\mu, i} \left[W_{x,i} \right]_\mathrm{ah} \right) },
\end{align}
where $\left[W_{x,i} \right]_\mathrm{ah}$ denotes the anti-Hermitian traceless part of $W_{x,i}$, and $\beta_{\mu, i} \in \mathbb R$ are real-valued weight parameters with $0 \le \mu \le D$ and $1 \le i \le N_\mathrm{ch}$. The above method projects $W_{x,i}$ onto the Lie algebra, and therefore $\mathcal{E}_{x,\mu}$ is guaranteed to be an element of the Lie group.

\textbf{Trace} layers generate gauge invariant output 
\begin{align}
    \mathcal{T}_{x,i}(\mathcal{U}, \mathcal{W}) &= \Tr \left[ W_{x,i} \right].
\end{align}

\textbf{Plaquette layers (\Plaq)} generate all possible plaquettes $U_{x,\mu\nu}$ from Eq.~(3) at location $x$ and add them to $\mathcal W$ as a preprocessing step. To reduce redundancy, we can choose to only compute plaquettes with positive orientation, i.e.~$U_{x,\mu\nu}$ with $\mu < \nu$. 

\textbf{Polyakov layers (\Poly)} compute all possible Polyakov loops~\cite{Polyakov:1978vu} at every lattice site according to
\begin{align}
    \mathcal{L}_{x,\mu}(\mathcal{U}) = \prod_k U_{x + k \cdot \mu,\mu} = U_{x,\mu} U_{x+\mu,\mu}  \dots  U_{x-\mu,\mu}
\end{align}
and add them to the set of locally transforming objects in $\mathcal W$ as a preprocessing step. These loops wrap around the periodic boundary of the (torus-like) space-time lattice and cannot be contracted to a single point. 

\begin{figure}
    \centering
    \includegraphics[width=0.45\textwidth]{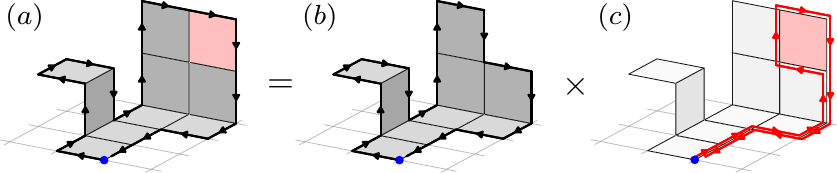}
    \caption{Sketch of the proof that \LGCNN{}s can generate arbitrary Wilson loops. 
    (a) An arbitrary contractible Wilson loop (depicted here in 3 dimensions) surrounds a surface that can be tessellated into $n$ tiles of \mbox{$1 \times 1$} unit lattice area. The blue dot indicates the starting point of the untraced Wilson loop.
    (b) A Wilson loop with $n$ tiles can be composed of an untraced Wilson loop with $n-1$ tiles and a path along the boundary to the missing tile using an \LBL{}.
    (c) An arbitrary return path to and from a \mbox{$1 \times 1$} plaquette is obtained by successive applications of \LConv{}s after an initial \Plaq{}.}
    \label{fig:proof}
\end{figure}

Figure~\ref{fig:proof} contains a sketch of the proof by induction that \LGCNN{}s can generate arbitrary Wilson loops (Fig.~\ref{fig:proof}a). This is achieved by concatenating loops as shown using an \LBL{} such that intermediate path segments to the origin (indicated by a blue dot in Fig.~\ref{fig:proof}) cancel. Arbitrary paths to a plaquette and back along the same path as shown in Fig.~\ref{fig:proof}c can be generated by an initial \Plaq{} with repeated application of \LConv{}s. On topologies that are not simply connected, loops that cannot be contracted to a point can be added by \Poly{}. The possibility of forming a complete Wilson loop basis~\cite{Giles:1981ej,Loll:1992fk} together with the universality of deep convolutional neural networks~\cite{Zhou:2018a} makes \LGCNN{}s capable of universal approximation within an equivalence class of gauge connections.

These layers can be assembled and applied to specific problems in lattice gauge theory. A possible architecture is depicted in Fig.~\ref{fig:layers}. The alternated application of \LConv{} and \LBL{} can double the area of loops. Repeating this block can grow Wilson loops to arbitrary size. \LBL{}s are already non-linear, but even more general relations can be expressed through \LAct{}s. Building blocks in the form of \LConv{}+\LBL{}+\LAct{} cover a wide range of possible gauge equivariant non-linear functions. The Trace layer renders the output gauge invariant so that it can be further processed by a conventional CNN or a multilayer perceptron (MLP) without spoiling gauge symmetry. Some applications, such as classical time evolution~\cite{Ambjorn:1990pu} or gradient flow~\cite{Luscher:2010iy}, require operations that can change the set of gauge links ~$\mathcal{U}$. This can be achieved using an \LExp{}. After an \LExp{}, one can use \Plaq{} and \Poly{} to update $\mathcal W$ accordingly.

We demonstrate the performance of \LGCNN{}s by applying them to a number of seemingly simple regression problems. Specifically, we train \LGCNN{} models using supervised learning to predict local, gauge invariant observables and make comparisons to traditional CNN models as a baseline test. We perform our experiments on data from 1+1D and 3+1D lattices with various sizes and coupling constants $g$, which we have generated using our own $\SU(2)$ Monte Carlo code based on the Metropolis algorithm~\cite{Creutz:1980bbb}. One type of observable that we focus on is the real value of traced Wilson loops, i.e.
\begin{align}
    W^{(m \times n)}_{x,\mu\nu} = \frac{1}{N_c} \mathrm{Re}\,  \Tr \left[ U^{(m \times n)}_{x,\mu\nu} \right],
\end{align}
where $U^{(m \times n)}_{x,\mu\nu}$ is an \mbox{$m\times n$} Wilson loop in the \mbox{$\mu\nu$} plane. A second observable that we study, which is of more immediate physical relevance, is the topological charge density $q_x$, which is only available in 3+1D. In particular, we focus on the plaquette discretization given by
\begin{align} \label{eq:top_charge_plaq}
    q^\mathrm{plaq}_x = \frac{\epsilon_{\mu\nu\rho\sigma}}{32 \pi^2}  \Tr \left[  \frac{U_{x,\mu\nu} \! - \! U^\dg_{x,\mu\nu}}{2i} \,  \frac{U_{x,\rho\sigma} \! - \! U^\dg_{x,\rho\sigma}}{2i}  \right].
\end{align}

Our frameworks of choice are \textit{PyTorch} and \textit{PyTorch Lightning}. We have implemented the necessary layers discussed previously as modules in \textit{PyTorch}, which can be used to assemble complete \LGCNN{} models. Our code is open source and hosted on GitLab \footnote{Our repository is hosted at \href{https://gitlab.com/openpixi/lge-cnn}{https://gitlab.com/openpixi/lge-cnn}}. In addition to gauge equivariance, we formulate our models to be translationally equivariant, which makes them applicable to arbitrary lattices. The task of the training procedure is to minimize a mean-squared error (MSE) loss function, which compares the prediction of the model to the ground truth from the dataset. For technical details, see our Supplementary Material.

Our \LGCNN{} architectures consist of stacks of \LConv{}+\LBL{} blocks, followed by a trace operation, as shown in Fig.~\ref{fig:layers}. The gauge invariant output at each lattice site is mapped by linear layers to the final output nodes. We have experimented with architectures of various sizes, with the smallest models only consisting of a single \LConv{}+\LBL{} layer and $\approx 100$ parameters to very large architectures with a stack of up to four layers of \LConv{}+\LBL{} and $\approx 40,000$ trainable parameters.

For comparison, we implement gauge symmetry breaking baseline models using a typical CNN architecture. We use stacks of two-dimensional convolutions followed by non-linear activation functions (such as ReLU, LeakyReLU, $\tanh{}$ and sigmoid) and global average pooling~\cite{Lin:2013gap} before mapping to the output nodes using linear layers. Baseline architectures vary from just one or two convolutions with \mbox{$\approx 300$} parameters to large models with up to six convolutions and \mbox{$\approx 100,000$} trainable weight parameters. These models are trained and validated on small lattices ($8 \cdot 8$ for 1+1D and $4 \cdot 8^3$ for 3+1D, $10^4$ training and $10^3$ validation examples) but tested on data from larger lattices (up to $64\cdot 64$ and $8 \cdot 16^3$, $10^3$ test examples). In total, we have trained $2680$ individual baseline models.

\begin{figure}
    \centering
    \includegraphics[width=0.45\textwidth]{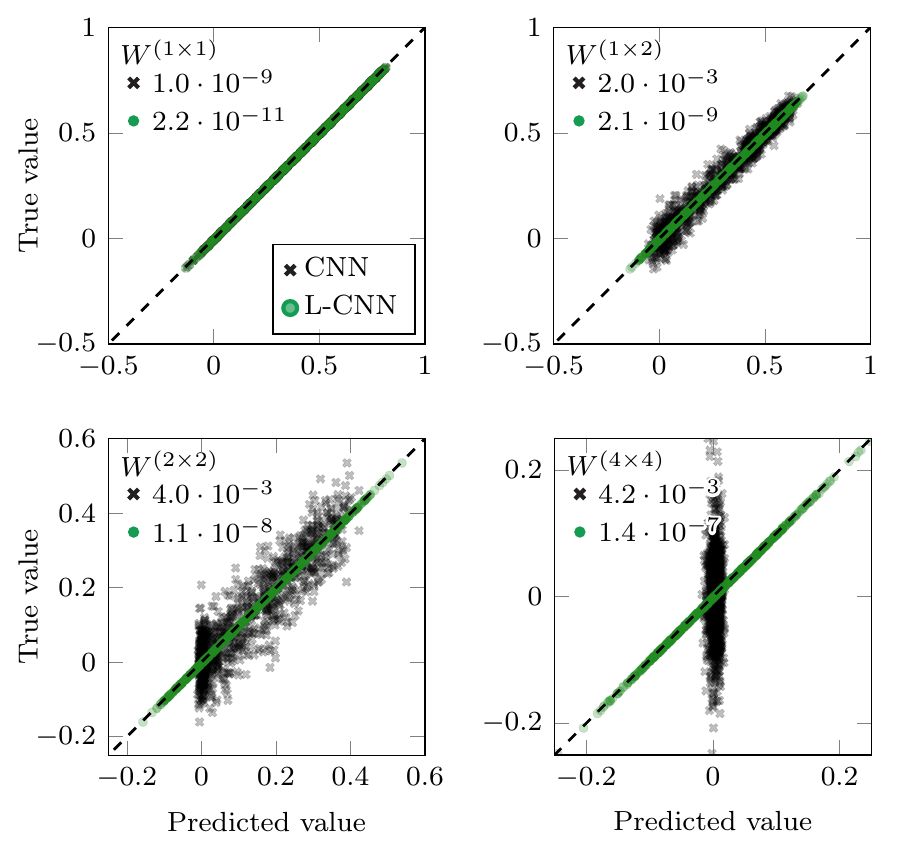}
    \caption{Scatter plots comparing best \LGCNN{} models to baseline CNN models for Wilson loops of various sizes for 1+1D. For each example in the $N_s \cdot N_t = 8 \cdot 8$ test dataset, we plot the true value vs.~the model prediction. Perfect agreement 
    is indicated by the dashed $45^\circ$ line. 
    As the size of the traced Wilson loops grows, the performance of the baseline CNN models worsens quickly. On the other hand, \LGCNN{} models achieve high agreement in all cases. The values in the upper left corner denote the MSEs of each plot.}
    \label{fig:d2_scatter}
\end{figure}

\begin{figure}
    \centering
    \includegraphics{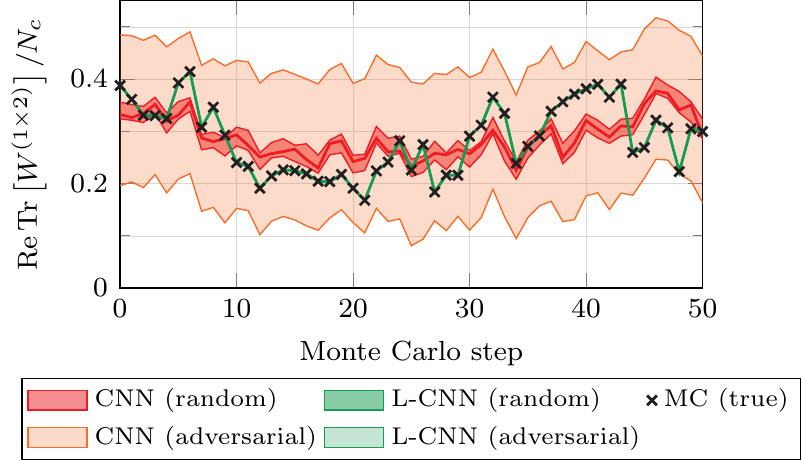}
    \caption{Prediction uncertainty in $W^{(1 \times 2)}$ due to breaking of gauge symmetry for our best baseline CNN and \LGCNN{} models on $8 \cdot 8$ test data. Black crosses (MC) denote the calculated true value of the Wilson loop. The red bands show the effects of random gauge transformations and transformations obtained from adversarial attacks. Predictions by the \LGCNN{} models are invariant by construction.} \label{fig:gauge_uncertainty}
\end{figure}

\begin{figure}
    \centering
    \includegraphics{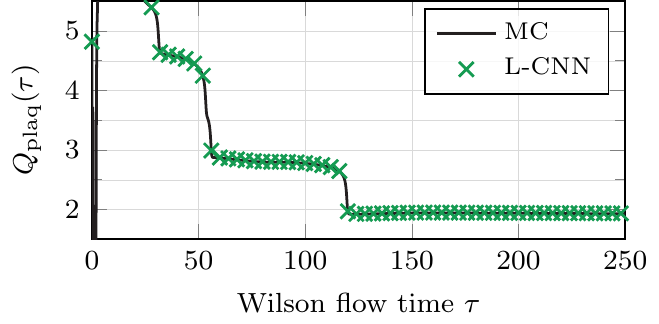}
    \caption{Predictions of our best \LGCNN{} for the topological charge on a Wilson flowed \mbox{$8 \cdot 24^3$} lattice configuration at $2 / g^2 = 0.2$.}
    \label{fig:qp_flow}
\end{figure}

A selection of results are presented in Figs.~\ref{fig:d2_scatter} and~\ref{fig:gauge_uncertainty} for 1+1D and Fig.~\ref{fig:qp_flow} for 3+1D lattices. Figure~\ref{fig:d2_scatter} shows scatter plots of our best performing models (\LGCNN{} and baseline) evaluated on test data. We demonstrate that the performance of the baseline models quickly deteriorates with the growing size of the Wilson loop. In the case of \mbox{$4 \times 4$} loops, the baseline model collapses and only predicts the average value of the training data. This signals that the baseline models are unable to learn any meaningful relationship between input and output data. Except for the case of \mbox{$1 \times 1$} loops, the baseline CNN models are not able to adequately learn even moderately sized Wilson loops in 1+1D and have particular difficulty to predict negative values, which are associated with large gauge rotations.

We have experimented with different baseline CNN architectures of various widths and depths and a variety of activation functions. In all of our experiments we have obtained similar behaviour as shown in Fig.~\ref{fig:d2_scatter}. In contrast, \LGCNN{} architectures are able to converge to solutions that can predict the observables to a high degree of accuracy in all tasks, and they are gauge covariant by construction. Furthermore, our models perform well across all considered lattice sizes because of translational equivariance. In Fig.~\ref{fig:gauge_uncertainty}, we show how predictions of baseline CNNs and \LGCNN{}s change under gauge transformations. We have tried two strategies: random gauge transformations and adversarial attacks (see our Supplementary Material Sec.~VI.A). We observe that trained CNN models learn to approximate gauge invariance (see~\cite{Boyda:2020nfh} for a similar result) but are vulnerable to certain gauge transformations that can drastically change their predictions.

\LGCNN{}s can also be applied to 3+1D lattices: Figure~\ref{fig:qp_flow} demonstrates the predictions of our best \LGCNN{} model for topological charge (trained on $4 \cdot 8^3$) during Wilson (gradient) flow~\cite{Luscher:2010iy} on an $8 \cdot 24^3$ lattice. The values from our simulation (MC) agree with the model predictions to high accuracy and assume integer values, as expected.

To summarize, we introduced a neural network structure for processing lattice gauge theory data that is capable of universal approximation within physically relevant degrees of freedom. The network achieves this by growing Wilson loops of arbitrary shapes in successive trainable gauge-covariant network layers. We demonstrated that our method surpasses ordinary convolutional networks in simple regression tasks on the lattice and that it manages to predict and generalize results for larger Wilson loops where a baseline network completely fails. Furthermore, our models can also be applied to lattices of any size without requiring retraining or transfer learning. From a broader perspective, we introduced a generalization of traditional CNNs that could replace them in a large range of machine learning applications where CNNs are applied to lattice gauge data.

Our approach opens up exciting possibilities for future research. So far, we implemented the network layers for the $\SU(2)$ gauge group only, but our method works for any $\SU(N_c)$. Also, we have introduced the general concepts of Polyakov-loop generating layers and of exponentiation layers, but we have not exploited them in numerical experiments. It would be interesting to study these layers and their possible applications. Finally, the compositional nature of successive gauge-covariant network layers is reminiscent of the renormalization group picture~\cite{Beny:2013a,Mehta:2014a,Li:2018a}. Trainable networks could provide a viable implementation of the renormalization group approaches by Wilson~\cite{Wilson:1980a} and Symanzik~\cite{Symanzik:1983dc}. Improved lattice actions and operators could be obtained by training on coarse lattices, while providing ground-truth data from finer grained simulations. 
Automatically learning improved lattice actions could make accessible previously unreachable system sizes for zero and finite temperature applications~\cite{Niedermayer:1996eb,Iwasaki:1985we,Lagae:1998pe} as well as for real-time lattice simulations~\cite{Moore:1996wn,Ipp:2018hai,Ipp:2017lho,Ipp:2020igo}.

\begin{acknowledgments}
DM thanks Jimmy Aronsson for valuable discussions regarding group equivariant and gauge equivariant neural networks. This work has been supported by the Austrian Science Fund FWF No.~P32446-N27, No.~P28352 and Doctoral program No. W1252-N27. The Titan\,V GPU used for this research was donated by the NVIDIA Corporation.
\end{acknowledgments}

\bibliography{references.bib}

\end{document}


\title{Supplementary Material: \\ Lattice gauge equivariant convolutional neural networks}

\author{Matteo Favoni}
\email{favoni@hep.itp.tuwien.ac.at}
\author{Andreas Ipp}
\email{ipp@hep.itp.tuwien.ac.at}
\author{David I.~M\"uller}
\email[Corresponding author: ]{dmueller@hep.itp.tuwien.ac.at}
\author{Daniel Schuh}
\email{schuh@hep.itp.tuwien.ac.at}
\affiliation{Institute for Theoretical Physics, TU Wien, Austria}

\date{\today}

\maketitle

In this Supplementary Material we present technical details related to our lattice gauge theory code and our implementation of \LGCNN{}s. We also describe how we performed the training for both \LGCNN{}s and baseline models.

\section{Monte Carlo simulation of pure lattice gauge theory}
We give a short review (largely following~\cite{Gattringer:2010abc}) of how to generate lattice gauge configurations using a Markov Chain  Monte Carlo (MCMC) for pure $\SU(2)$ lattice gauge theory.

We use the Wilson action
\begin{align}
    S_W[U] &= \frac{\beta}{N_c} \sum_{x \in \Omega}\sum_{\mu < \nu} \Re \Tr \left[ \one - U_{x,\mu\nu}\right],
\end{align}
where the coupling constant $\beta$ is related to the Yang-Mills coupling $g$ via $\beta = 2 N_c / g^2$. The number of colors is  $N_c = 2$ and plaquettes are defined as
\begin{align}
    U_{x,\mu\nu} &= U_{x,\mu} U_{x+\mu, \nu} U^\dg_{x+\nu, \mu} U^\dg_{x,\nu},
\end{align}
with $U_{x,\mu} \in \SU(N_c)$ denoting a gauge link connecting the lattice site $x \in \Lambda$ to $x+\mu \in \Lambda$. The Wilson action is invariant under gauge transformations $\Omega_x \in \SU(N_c)$
\begin{align}
    U_{x,\mu} \rightarrow \Omega_x U_{x,\mu} \Omega^\dg_{x+\mu}.
\end{align}

In order to generate sequences of random configurations $\mathcal U$ according to the probability functional
\begin{align}
    \rho[U] \propto e^{-S_W[U]},
\end{align}
we propose random updates of the form
\begin{align}
    U'_{x,\mu} = V U_{x,\mu}, \quad V \in \SU(N_c), \label{eq:update}
\end{align}
where $V$ is a random $\SU(N_c)$ matrix that is close to the unit element and accept or reject them according to the update probability
\begin{align}
    p[U, U'] = \mathrm{min}\left(1, e^{-S_W[U'] + S_W[U]} \right).
\end{align}
Random update matrices $V$ are generated by first generating random color vectors $X^a = A\, \eta^a$ with amplitude $A > 0$ and $a \in \{ 1, 2, \dots, N_c^2 - 1 \}$, where $\eta^a$ are uncorrelated standard random normal variables. Using matrix exponentiation and the generators of the gauge group $T^a$, we can write the random matrix $V$ as
\begin{align}
    V = e^{i \sum_a T^a X^a}.
\end{align}

We perform multiple updates of a single link consecutively. A whole sweep consists of performing these updates for every link in the configuration. For each sweep we use the amplitude $A = 0.5$ and update each link ten times. 

Our implementation of this method has been written in \textit{Python} and \textit{Numba} and is included with the \textit{lge-cnn} repository \bibnotemark[repo].
\bibnotetext[repo]{Our repository is hosted at \href{https://gitlab.com/openpixi/lge-cnn}{https://gitlab.com/openpixi/lge-cnn}}

\section{Datasets \label{sec:datasets}}
\renewcommand*\arraystretch{1.4}
\begin{table}
    \caption{\label{tab:datasets} Details about the datasets for 1+1D and 3+1D lattices. Test sets contain $10^3$ examples for each individual lattice size.}
    \begin{ruledtabular}
        \begin{tabular}{p{20mm} | l | l | p{25mm}}
            \textbf{1+1D} & & & \\
            \hline
            &Training & Validation &  Test\\
            \hline
            Lattice $N_t \cdot N_s$& $8 \! \cdot \! 8$ & $8 \! \cdot \! 8$ & ${8 \! \cdot \! 8}$, ${16 \! \cdot \! 16}$, ${32 \! \cdot \! 32}$, ${64 \! \cdot \! 64}$\\ 
            \hline
            Examples & $10^4$ & $10^3$ & $10^3$ per lattice \\
            \hline
            Labels & \multicolumn{3}{l}{$W^{(1 \times 1)}_{x,01}$, $W^{(1 \times 2)}_{x,01}$,  $W^{(2 \times 2)}_{x,01}$, $W^{(4 \times 4)}_{x,01}$ } \tabularnewline
            \hline
            Coupling  & \multicolumn{3}{l}{$\beta \in \{ 0.1, \dots, 6.0\}$} \tabularnewline
            \hline
            \hline
            \textbf{3+1D} & & & \\
            \hline
            &  Training &  Validation & Test\\
            \hline
            Lattice $N_t \cdot N_s^3$& $4 \! \cdot \! 8^3$ & $4 \! \cdot \!  8^3$ & ${4 \! \cdot \! 8^3}$, ${6 \! \cdot \! 8^3}$, ${6 \! \cdot \! 12^3}$, ${8 \! \cdot \! 16^3}$\\ 
            \hline
            Examples & $10^4$ & $10^3$ & $10^3$ per lattice  \\
            \hline
            Labels & \multicolumn{3}{l}{$W^{(1 \times 1)}_{x,12}$, $W^{(1 \times 2)}_{x,12}$, $W^{(2 \times 2)}_{x,12}$, $W^{(4 \times 4)}_{x,12}$, $q^\mathrm{plaq}_x$ } \tabularnewline
            \hline Coupling  & \multicolumn{3}{l}{$\beta \in \{ 0.1, \dots, 6.0\}$} \tabularnewline
        \end{tabular}
    \end{ruledtabular}
\end{table}

\renewcommand*\arraystretch{1.0}
In this section we present details about the datasets that we use to train both baseline models and \LGCNN{} models. All datasets are summarized in table~\ref{tab:datasets}. 

Each example of a particular dataset consists of a tuple $(\mathcal{U}, \mathcal{W})$ and a few gauge invariant observables (or labels) $\{ \mathcal{O}_i \}$. Here, $\mathcal{U}$ denotes the set of gauge links, and $\mathcal{W}$ stands for all \mbox{$1 \times 1$} Wilson loops $U_{x,\mu\nu}$ with positive orientation (in the first orthant $0\leq \mu < \nu$) that can be computed from the set of gauge links. In the case of the gauge group $\SU(N_c)$ on a $N_t \cdot N_s^{D}$ lattice, the tuple $(\mathcal{U}, \mathcal{W})$ can be represented by an array of 
\begin{align}
    N_\mathrm{input} = 2 \, N^2_c \cdot N_t \cdot N_s^D \cdot ((D+1) + (D+1) D / 2 ) \label{eq:dataset_input_size}
\end{align}
real numbers. Due to the scaling with $D$, datasets can quickly grow in size. For this reason we use the \textit{HDF5} file format for storage.

The samples of our datasets are generated by the previously described MCMC method. After a random initialization similar to the random update defined in Eq.~\eqref{eq:update}, we perform $N_\mathrm{warmup} = 2 \cdot 10^3$ sweeps to thermalize the system. Afterwards, every $N_\mathrm{obs} = 10^2$ sweeps we save the tuple $(\mathcal{U}, \mathcal{W})$ and compute the desired observables $\{ \mathcal{O}_i \}$. The coupling constant $\beta$  varies from $\beta_\mathrm{min} = 0.1$ to $\beta_\mathrm{max} = 6.0$ in equal steps of $\Delta \beta = (\beta_\mathrm{max} - \beta_\mathrm{min}) / N_\beta$ with $N_\beta = 10$. We have created datasets for 1+1D and 3+1D lattices. In both cases we have generated training ($10^4$ total examples) and validation sets ($10^3$ total examples) only for the smallest lattice sizes. Test sets ($10^3$ total examples) were created for larger lattice sizes as well. Observables (or labels) differ depending on the dimension of the lattice. In 1+1D and 3+1D we compute (real) traces of \mbox{$1 \times 1$}, \mbox{$1 \times 2$}, \mbox{$2 \times 2$} and \mbox{$4 \times 4$} Wilson loops given  by
\begin{align}
    W^{(m \times n)}_{x,\mu\nu} = \frac{1}{N_c} \mathrm{Re}\,  \Tr \left[ U^{(m \times n)}_{x,\mu\nu} \right],
\end{align}
where $U^{(m \times n)}_{x,\mu\nu}$ is an \mbox{$m \times n$} Wilson loop in the $\mu\nu$ plane starting at $x$. In 3+1D we restrict ourselves to Wilson loops in the $xy$ plane, whereas in 1+1D there is only one possible plane. Additionally, we compute the local topological charge density $q(x)$ in 3+1D using the plaquette definition~\cite{Alexandrou:2017hqw}
\begin{align}
    q^\mathrm{plaq}_x = \frac{\epsilon_{\mu\nu\rho\sigma}}{32 \pi^2}  \Tr \left[  \frac{U_{x,\mu\nu} \! - \! U^\dg_{x,\mu\nu}}{2i} \,  \frac{U_{x,\rho\sigma} \! - \! U^\dg_{x,\rho\sigma}}{2i}  \right]. \label{eq:top_charge_plaq_supp}
\end{align}

\section{Implementation details} \label{sec:implementation}
The gauge equivariant operations defined in this work can be realized in any modern machine learning framework. For this work we chose \textit{PyTorch}, although an implementation in \textit{TensorFlow} should be straightforward as well. The code for our proof-of-principle implementation is hosted in a public GitLab repository~\citenote{repo}.

During the development of the code, \textit{PyTorch} did not fully support operations on complex numbers. For this reason, our code uses a split of complex matrices into real and imaginary parts. Complex matrix multiplication is reduced to real matrix multiplications of real and imaginary parts. In practice, we perform batched matrix multiplication using the \textit{torch.einsum} method of \textit{PyTorch}. For simplicity, we assumed weight parameters to be real-valued. Gauge links and locally transforming matrices (such as plaquettes) are stored and processed as \textit{PyTorch} \textit{tensors}. In order to support lattices of arbitrary size and dimension using the same code base, the $(D+1)$-dimensional lattice is flattened out by default. This means that the dimension of the \textit{tensor} is always the same, irrespective of the actual dimension of the lattice. Only when performing translations, as required e.g.~in a convolution, the lattice structure is restored using \textit{torch.reshape} or \textit{torch.Tensor.view}. The reshaped \textit{tensor} can then be shifted using \textit{torch.roll} along a particular axis and is flattened out again afterwards. This allows us to perform convolutions in any lattice dimension. At the cost of some computational overhead, this allows our framework to be highly flexible and simple to maintain. It should be noted that these technical details are likely to change in future versions of our code.

In our implementation of the \LGCNN{} framework we combined lattice gauge equivariant convolutions (\LConv{}) and lattice gauge equivariant bilinear layers (\LBL{}) into a single module called \LCB{}. In practice, we first compute all parallel-transported terms
\begin{align}
    W'_{x+k \cdot \mu,j} =  U_{x,k\cdot \mu} W_{x+k \cdot \mu,j} U^\dg_{x,k\cdot \mu}, 
\end{align}
with $-K \leq k \leq K$ for kernel size $K$, which are then multiplied with all local terms $W_{x,j}$ in a bilinear layer
\begin{align}
    W_{x,i} \rightarrow \sum_{j,j',k} \alpha_{i, j, j', k} W_{x,j} W'_{x+k \cdot \mu,j'}.\label{eq:lcb}
\end{align}
In the above bilinear operation, we also include all Hermitian conjugates of both local terms $W$ and transported terms $W'$, as well as the unit element $\one$. This makes the \LCB{} operation highly flexible as it can also act as a residual module. We note that the \LCB{} operation is equivalent to a composition of \LConv{} and \LBL{} except that trainable weights are parametrized in a slightly different way. Empirically, we found that training models using this parametrization tends to be easier. Another simplification that we use by default is that we restrict our convolutions to only consider positive shifts along the lattice axes, i.e.~we use $0 \leq k \leq K$ unless explicitly specified otherwise.

Finally, in order to more easily train \LGCNN{} models on observables which tend to assume rather small numerical values, such as $q^\mathrm{plaq}_x$ given in Eq.~\eqref{eq:top_charge_plaq_supp}, we allow for a fixed scaling factor of the labels in our datasets:
\begin{align}
    \tilde q^\mathrm{plaq}_x = C q^\mathrm{plaq}_x, \qquad C > 0.
\end{align}
This is similar to simple ``whitening'' transformations used in generic machine learning applications, which normalize the domains of the output and input data. In our work we used $C=100$ for the regression task on $q^\mathrm{plaq}_x$, and $C=1$ for all Wilson loop regressions.

\section{Baseline networks}
\begin{table}
    \caption{Baseline CNN architectures for $W^{(1\times 1)}$ and $W^{(1\times 2)}$ regression tasks. The activation functions \textit{LeakyReLU}, \textit{ReLU}, \textit{sigmoid} and \textit{tanh} have been used for each of the architectures individually. No activation function has been applied to the output. For $W^{(1 \times 1)}$ the number of input channels $N_\mathrm{in}$ differs depending on the input data given to the CNN:
    $N_\mathrm{in} = 16$ for input consisting only of gauge links ($\mathcal{U}$),
    $N_\mathrm{in} = 24$ for gauge links and plaquettes with positive orientation ($\mathcal{U}, \mathcal{W}$),
    and $N_\mathrm{in} = 32$ for gauge links and plaquettes with both orientation ($\mathcal{U}, \mathcal{W}, \mathcal{W}^\dagger$). The number of trainable parameters is given by $N_\mathrm{param}$ and depends on the number of input channels.}
    \label{tab:arch_base_w1x2}
    \scriptsize
    \begin{ruledtabular}
        \begin{tabular}{l | l | l | l}
            \multicolumn{2}{l|}{$W^{(1 \times 1)}$, $W^{(1 \times 2)}$} & & \\
            \hline
            \hline
            Small & Architecture 1 & Architecture 2 & Architecture 3 \\
            \hline
            & \CCD{2}{$N_\mathrm{in}$}{4} & \CCD{2}{$N_\mathrm{in}$}{4} & \CCD{1}{$N_\mathrm{in}$}{8} \\
            & \CCD{1}{4}{8} & \CCD{2}{4}{4} & \CCD{2}{8}{4} \\
            & GAP & GAP & GAP \\
            & \Lin{8}{4} & \Lin{4}{4} & \Lin{4}{1} \\
            & \Lin{4}{1} & \Lin{4}{1} & - \\
            \hline
            $N_\mathrm{param}^{(U)^{\phantom{\dagger}}}$ & 341 & 353 & 273 \\
            \hline
            $N_\mathrm{param}^{(U, W)^{\phantom{\dagger}}}$ & 469 & 481 & 337 \\
            \hline
            $N_\mathrm{param}^{(U, W, W^\dagger)}$ & 597 & 609 & 401 \\
            \hline
            \hline
            Medium & Architecture 1 & Architecture 2 & Architecture 3 \\
            \hline
            & \CCD{2}{$N_\mathrm{in}$}{8} & \CCD{2}{$N_\mathrm{in}$}{8} & \CCD{3}{$N_\mathrm{in}$}{4} \\
            & \CCD{2}{8}{8} & \CCD{2}{8}{8} & \CCD{2}{4}{8} \\
            & \CCD{2}{8}{8} & - & - \\
            & GAP & GAP & GAP \\
            & \Lin{8}{4} & \Lin{8}{4} & \Lin{8}{4} \\
            & \Lin{4}{1} & \Lin{4}{1} & \Lin{4}{1} \\
            \hline
            $N_\mathrm{param}^{(U)^{\phantom{\dagger}}}$ & 1,089 & 825 & 757 \\
            \hline
            $N_\mathrm{param}^{(U, W)^{\phantom{\dagger}}}$ & 1,345 & 1,081 & 1,045 \\
            \hline
            $N_\mathrm{param}^{(U, W, W^\dagger)}$ & 1,601 & 1,337 & 1,333 \\
            \hline
            \hline
            Large & Architecture 1 & Architecture 2 & Architecture 3 \\
            \hline
            & \CCD{2}{$N_\mathrm{in}$}{16} & \CCD{3}{$N_\mathrm{in}$}{16} & \CCD{3}{$N_\mathrm{in}$}{16} \\
            & \CCD{2}{16}{16} & \CCD{3}{16}{8} & \CCD{1}{16}{8} \\
            & \CCD{2}{16}{16} & - & \CCD{3}{8}{16} \\
            & GAP & GAP & GAP \\
            & \Lin{16}{8} & \Lin{8}{8} & \Lin{16}{8} \\
            & \Lin{8}{1} & \Lin{8}{1} & \Lin{8}{1} \\
            \hline
            $N_\mathrm{param}^{(U)^{\phantom{\dagger}}}$ & 3,265 & 3,561 & 3,769 \\
            \hline
            $N_\mathrm{param}^{(U, W)^{\phantom{\dagger}}}$ & 3,777 & 4,713 & 4,921 \\
            \hline
            $N_\mathrm{param}^{(U, W, W^\dagger)}$ & 4,289 & 5,865 & 6,073 \\
            \hline
            \hline
            Wide & Architecture 1 & Architecture 2 & Architecture 3 \\
            \hline
            & \CCD{2}{$N_\mathrm{in}$}{128} & \CCD{2}{$N_\mathrm{in}$}{256} & \CCD{2}{$N_\mathrm{in}$}{512} \\
            & - & \CCD{3}{256}{32} & - \\
            & GAP & GAP & GAP \\
            & \Lin{128}{1} & \Lin{32}{1} & \Lin{512}{64} \\
            & - & - & \Lin{64}{1} \\
           \hline
            $N_\mathrm{param}^{(U)^{\phantom{\dagger}}}$ & 8,449 & 90,433 & 66,177 \\
            \hline
            $N_\mathrm{param}^{(U, W)^{\phantom{\dagger}}}$ & 12,545 & 98,625 & 82,561 \\
            \hline
            $N_\mathrm{param}^{(U, W, W^\dagger)}$ & 16,641 & 106,817 & 98,945 \\
        \end{tabular}
    \end{ruledtabular}
\end{table}

\begin{table}
    \caption{Baseline CNN architectures for $W^{(2\times 2)}$ regression tasks. We use the same notation as in table~\ref{tab:arch_base_w1x2}. \label{tab:arch_base_w2x2}}
    \scriptsize
    \begin{ruledtabular}
        \begin{tabular}{l | l | l | l}
            $W^{(2 \times 2)}$ & & & \\
            \hline
            \hline
            Small  & Architecture 1 \hspace{2em} & Architecture 2 \hspace{2em} & Architecture 3 \hspace{2em}  \\
            \hline
            & \CCD{2}{32}{4} & \CCD{2}{32}{2} & \CCD{2}{32}{4}  \\
            & \CCD{2}{4}{4} & \CCD{1}{2}{4} & \CCD{2}{4}{2} \\
            & GAP & GAP & GAP \\
            & \Lin{4}{4} & \Lin{4}{1} & \Lin{2}{1} \\
            & \Lin{4}{1} & - &  \\
            \hline
            $N_\mathrm{param}$ & 609 & 275 & 553 \\
            \hline
            \hline
            Medium & Architecture 1 & Architecture 2 & Architecture 3 \\
            \hline
            & \CCD{2}{32}{4} & \CCD{2}{32}{8} & \CCD{3}{32}{4} \\
            & \CCD{2}{4}{8} & \CCD{2}{8}{8} & \CCD{2}{4}{8} \\
            & \CCD{2}{8}{8} & \CCD{2}{8}{8} & \CCD{3}{8}{8} \\
            & \CCD{2}{8}{8} & \CCD{2}{8}{8} & \CCD{2}{8}{8} \\
            & GAP & GAP & GAP \\
            & \Lin{8}{16} & \Lin{8}{8} & \Lin{8}{4} \\
            & \Lin{16}{1} & \Lin{8}{1} & \Lin{4}{1} \\
            \hline
            $N_\mathrm{param}$ & 1,341 & 1,905 & 2,181 \\
            \hline
            \hline
            Large & Architecture 1 & Architecture 2 & Architecture 3 \\
            \hline
            & \CCD{2}{32}{8} & \CCD{2}{32}{8} & \CCD{3}{32}{8} \\
            & \CCD{2}{8}{16} & \CCD{2}{8}{16} & \CCD{3}{8}{16} \\
            & \CCD{2}{16}{32} & \CCD{2}{16}{32} & \CCD{3}{16}{32} \\
            & \CCD{2}{32}{64} & \CCD{2}{32}{64} & \CCD{3}{32}{16} \\
            & - & \CCD{2}{64}{32} & - \\
            & GAP & GAP & GAP \\
            & \Lin{64}{16} & \Lin{32}{8} & \Lin{16}{8} \\
            & \Lin{16}{1} & \Lin{8}{1} & \Lin{8}{1} \\
            \hline
            $N_\mathrm{param}$ & 12,953 & 20,393 & 12,889
        \end{tabular}
    \end{ruledtabular}
\end{table}

\begin{table}
    \caption{Baseline CNN architectures for $W^{(4\times 4)}$ regression tasks. We use the same notation as in table~\ref{tab:arch_base_w1x2}. \label{tab:arch_base_w4x4}}
    \scriptsize
    \begin{ruledtabular}
        \begin{tabular}{l | l | l | l}
            $W^{(4 \times 4)}$ & & & \\
            \hline
            \hline
            Small & Architecture 1 \hspace{2em} & Architecture 2 \hspace{2em} & Architecture 3 \hspace{2em} \\
            \hline
            & \CCD{2}{32}{4} & \CCD{2}{32}{4} & \CCD{2}{32}{4} \\
            & \CCD{2}{4}{4} & \CCD{1}{4}{8} & \CCD{2}{4}{2} \\
            & GAP & GAP & GAP \\
            & \Lin{4}{4} & \Lin{8}{4} & \Lin{2}{1} \\
            & \Lin{4}{1} & \Lin{4}{1} & - \\
            \hline
            $N_\mathrm{param}$ & 609 & 597 & 553 \\
            \hline
            \hline
            Medium & Architecture 1 & Architecture 2 & Architecture 3 \\
            \hline
            & \CCD{3}{32}{16} & \CCD{2}{32}{16} & \CCD{3}{32}{8} \\
            & \CCD{1}{16}{8} & \CCD{2}{16}{24} & \CCD{2}{8}{16} \\
            & \CCD{3}{8}{16} & \CCD{2}{24}{16} & \CCD{1}{16}{32} \\
            & - & - & \CCD{2}{32}{16} \\
            & - & - & \CCD{2}{16}{8} \\
            & GAP & GAP & GAP \\
            & \Lin{16}{8} & \Lin{16}{8} & \Lin{8}{8} \\
            & \Lin{8}{1} & \Lin{8}{1} & \Lin{8}{1} \\
            \hline
            $N_\mathrm{param}$ & 6,073 & 5,321 & 6,049 \\
            \hline
            \hline
            Large & Architecture 1 & Architecture 2 & Architecture 3 \\
            \hline
            & \CCD{3}{32}{16} & \CCD{2}{32}{16} & \CCD{4}{32}{16} \\
            & \CCD{3}{16}{32} & \CCD{2}{16}{32} & \CCD{4}{16}{32} \\
            & \CCD{3}{32}{64} & \CCD{2}{32}{64} & \CCD{4}{32}{32} \\
            & \CCD{3}{64}{32} & \CCD{2}{64}{64} & \CCD{4}{32}{16} \\
            & - & \CCD{2}{64}{32} & - \\
            & - & \CCD{2}{32}{16} & - \\
            & GAP & GAP & GAP \\
            & \Lin{32}{16} & \Lin{16}{16} & \Lin{16}{8} \\
            & \Lin{16}{1} & \Lin{16}{8} & \Lin{8}{8} \\
            & - & \Lin{8}{1} & \Lin{8}{1} \\
            \hline
            $N_\mathrm{param}$ & 46,769 & 39,553 & 41,273
        \end{tabular}
    \end{ruledtabular}
\end{table}

The baseline models that we use in our comparison study are translationally equivariant deep convolutional neural networks of various widths and depths. We focus on networks with translational equivariance because pure lattice gauge theory is symmetric under lattice translations. Equivariance is guaranteed by choosing architectures consisting solely of convolutional layers that use stride parameters of one with periodic boundary conditions and by using a global average pooling layer (GAP) after the convolutional part of the network~\cite{Bulusu:2021rqz}. As a result of this architecture choice, a translation of the inputs would leave the predictions of our baseline networks invariant. Requiring translational equivariance forces us to use $D$-dimensional convolutions for $D$-dimensional lattices. At the time of writing, our machine learning framework of choice, \textit{PyTorch}, ships with implementations for one-, two- and three-dimensional convolutions. We are therefore restricted to lower dimensional lattices in the case of baseline architectures.

Tables~\ref{tab:arch_base_w1x2}, \ref{tab:arch_base_w2x2} and~\ref{tab:arch_base_w4x4} summarize the details of our baseline architectures. Our intent was to cover a large range of different architectures: both shallow and deep, but also wide (i.e.~a large number of channels) networks were tried. For each type (small, medium, large, wide), we manually picked three arbitrary architectures of roughly similar size.

In these tables we denote two-dimensional convolutions with periodic boundary conditions (realized by ``circular padding'') by $\CCD{K}{N_\mathrm{in}}{N_\mathrm{out}}$, where $K$ denotes the size of the quadratic kernel \mbox{$(K \times K)$}, and $N_\mathrm{in}$ and $N_\mathrm{out}$ are the number of input and output channels. ``GAP'' denotes a global average pooling layer and \Lin{$N_\mathrm{in}$}{$N_\mathrm{out}$} is a fully-connected linear (or dense) layer with $N_\mathrm{in}$ input and $N_\mathrm{out}$ output nodes. Non-linear activation functions are applied after every convolution and linear layer, except the final output layer. For each architecture, we try out four different activation functions: \textit{LeakyReLU}, \textit{ReLU}, \textit{sigmoid} and \textit{tanh}. We also list the number of trainable parameters $N_\mathrm{param}$. 

In order to feed the lattice gauge field configurations $\mathcal{U}$ to a CNN, the data has to be shaped into an appropriate format. Since we are working with neural networks that need real-valued input, the complex matrix elements of gauge links have to be split into real and imaginary values. More specifically, we ``flatten'' gauge links $U$ in the form of complex \mbox{$N_c \times N_c$} matrices into real-valued vectors of length $2 \cdot N^2_c$. For example, in the case of $N_c = 2$ we use
\begin{equation}
    U = \begin{pmatrix}
    a & b \\
    c & d
    \end{pmatrix} \rightarrow \left( \mathrm{Re} \, a, \mathrm{Re} \, b, \mathrm{Re} \, c, \mathrm{Re} \, d, \mathrm{Im} \, a, \cdots \right).
\end{equation}
Each link is therefore represented by $2 \cdot N_c^2 = 8$ real-valued numbers. Thus, a gauge field configuration in 1+1D on an $N_t \cdot N_s$ lattice with two links per lattice site can be represented by a real-valued \textit{PyTorch} tensor with shape $(N_\mathrm{in}, N_t, N_s)$, where we have $N_\mathrm{in} = 16$ channels. While gauge links $\mathcal{U}$ alone are in principle sufficient as an input, a fair comparison to \LGCNN{}s is only possible if baseline networks are also provided with pre-computed plaquettes $\mathcal{W}$ in the input layer. We show this in the case of $W^{(1 \times 1)}$ loops, where we carry out computational experiments regarding the performance of baseline networks based on the exact input the networks are provided with.  More specifically, we compare three cases: only gauge links $\mathcal{U}$ as input, gauge links and plaquettes with positive orientation $(\mathcal{U}, \mathcal{W})$, and gauge links and plaquettes of both orientations $(\mathcal{U}, \mathcal{W}, \mathcal{W}^\dg)$. Depending on the type of input, the number of input channels shown in table~\ref{tab:arch_base_w1x2} differs. If the models are provided with links $\mathcal{U}$, we have $N_\mathrm{in} = 2 \cdot 2 \cdot N_c^2 = 16$ real-valued input channels. For $(\mathcal{U}, \mathcal{W})$ we add eight additional real-valued channels, and for $(\mathcal{U}, \mathcal{W}, \mathcal{W}^\dg)$ we add sixteen, leading to a total of $32$ input channels. Therefore, the number of trainable parameters depends on the type of input. We list them in table~\ref{tab:arch_base_w1x2} for all three cases.  The results of this comparison is presented in section~\ref{sec:results}, where it is evident that $(\mathcal{U}, \mathcal{W})$ and $(\mathcal{U}, \mathcal{W}, \mathcal{W}^\dagger)$ lead to much better performance by many orders of magnitude. The input for baseline networks for larger Wilson loops, $W^{(1 \times 2)}$, $W^{(2 \times 2)}$ and $W^{(4 \times 4)}$, is consequently fixed to $(\mathcal{U}, \mathcal{W}, \mathcal{W}^\dagger)$. In summary, table~\ref{tab:arch_base_w1x2} lists 144 different architectures for $W^{(1 \times 1)}$ and 48 different architectures for $W^{(1 \times 2)}$, while tables~\ref{tab:arch_base_w2x2} and~\ref{tab:arch_base_w4x4} each list 36 different architectures.

Training of our baseline models was performed on an \textit{NVidia RTX 2070 Super} graphics card with eight gigabytes of memory using \textit{PyTorch} and \textit{PyTorch Lightning}. For each individual baseline architecture, ten independent models are trained to generate a model ensemble for each type of architecture. This allows us to average over random weight initializations and the stochastic nature of the optimization process. In particular, we use the \textit{AdamW} optimizer with a learning rate of $3\cdot 10^{-2}$ (zero weight decay) and train for up to 100 epochs using a batch size of 50. Early stopping (patience value of 25) based on validation loss is used to terminate training after a plateau is reached. In total, we trained $2680$ baseline models (including 40 models shown in table~\ref{tab:results_nogap}). On an \textit{NVidia Titan V} GPU both our smallest and largest baseline models require roughly 11 seconds per epoch (which indicates a bottleneck in terms of memory bandwidth during training) and one to two gigabytes of GPU memory. A single baseline model can be trained in less than 20 minutes.

We note that our baseline CNN architectures are very similar to those used in~\cite{Boyda:2020nfh}, where traditional CNNs were applied to regression tasks for the Polyakov loop for both SU(2) and SU(3) lattice configurations. Similarly, the authors of~\cite{Wetzel:2017ooo} use CNNs to detect the phase transition in SU(2) lattice gauge theory. We use the same basic established building blocks (convolutions, similar activation functions such as \textit{ReLU} and global averaging) and employ a similar training procedure (\textit{AdamW}). However, there are a few differences to our approach. The authors of~\cite{Boyda:2020nfh} use three-dimensional convolutions applied to four-dimensional lattice data. They achieve this by combining two of the spatial dimensions, effectively reducing the input dimension to three. As evident from their results, this works in the case of Polyakov loops along the temporal axis, which are unaffected by this flattening (or reshaping) operation. However, reshaping the input in this way leads to explicit breaking of spatial translational symmetry. For this reason we avoid using four-dimensional lattices in our baseline study and concentrate on two-dimensional data which can be properly processed with two-dimensional convolutions. By accounting for translational equivariance, the predictions of our baseline models are insensitive to space-time translations of the input data. We have found~\cite{Bulusu:2021rqz} that CNN architectures with exact translational equivariance led to much better results in regression tasks compared to symmetry-breaking models, although it should be noted that this study was carried out in the context of scalar field theory.  

\begin{table}[t]
    \caption{\LGCNN{} architectures for $W^{(1 \times 1)}$, $W^{(1 \times 2)}$, $W^{(2 \times 2)}$ and $W^{(4 \times 4)}$ in 1+1D. $N_\mathrm{param}$ denotes the number of trainable parameters. After the trace, the single linear layers are applied to each lattice site individually. No activation function is applied to the output. In contrast to our baseline architectures detailed in tables~\ref{tab:arch_base_w1x2}, \ref{tab:arch_base_w2x2} and~\ref{tab:arch_base_w4x4}, the \LGCNN{} models do not use a global average pooling layer. The output of the \LGCNN{} therefore consists of predictions for every lattice site. \label{tab:arch_lcnn_2d}}
    \scriptsize
    \begin{ruledtabular}
        \begin{tabular}{l | l | l | l}
            $W^{(1\times 1)}$ & & & \\
            \hline
            \hline
            & Small \hspace{5em} & \hspace{8em}  & \hspace{12em} \\
             \hline
            & \Conv{1}{1}{1} &  &  \\
            & \GTr &  &  \\
            & \Lin{2}{1} & & \\
             \hline
             $N_\mathrm{param}$ & 12 &  &  \\
             \hline
             \hline
            $W^{(1\times 2)}$ & & & \\
            \hline
            \hline
            & Small  & Medium & Large \\
             \hline
            & \Conv{2}{1}{2} & \Conv{3}{1}{4} & \Conv{4}{1}{8} \\
            & \GTr & \GTr & \GTr \\
            & \Lin{4}{1} & \Lin{8}{1} & \Lin{16}{1} \\
             \hline
             $N_\mathrm{param}$ & 35 & 117 & 329 \\
             \hline
             \hline
            $W^{(2\times 2)}$ & & & \\
            \hline
            \hline
            & Small & Medium & Large \\
             \hline
            & \Conv{2}{1}{2} & \Conv{3}{1}{4} & \Conv{4}{1}{8} \\
            & \Conv{2}{2}{2} & \Conv{3}{4}{4} & \Conv{4}{8}{8} \\
            & \GTr & \GTr & \GTr \\
            & \Lin{4}{1} & \Lin{8}{1} & \Lin{16}{1} \\
             \hline
             $N_\mathrm{param}$ & 125 & 1,305 & 13,521 \\
             \hline
             \hline
             $W^{(4 \times 4)}$ & & & \\
             \hline
             \hline
             & Small & Medium & Large \\
             \hline
            & \Conv{2}{1}{2} & \Conv{3}{1}{4} & \Conv{4}{1}{8} \\
            & \Conv{2}{2}{2} & \Conv{3}{4}{4} & \Conv{4}{8}{8} \\
            & \Conv{3}{2}{2} & \Conv{4}{4}{4} & \Conv{4}{8}{8} \\
            & \Conv{3}{2}{2} & \Conv{4}{4}{4} & \Conv{4}{8}{8} \\
            & \GTr & \GTr & \GTr \\
            & \Lin{4}{1} & \Lin{8}{1} & \Lin{16}{1} \\
             \hline
            $N_\mathrm{param}$ & 465 & 4,833 & 39,905 \\
        \end{tabular}
    \end{ruledtabular}
\end{table}

\section{\LGCNN{} networks}\label{sec:lcnn-networks}
\begin{table}
    \caption{\LGCNN{} architectures for $W^{(2 \times 2)}$,  $W^{(4 \times 4)}$ and $q^\mathrm{plaq}$ in 3+1D using the same notation as in table~\ref{tab:arch_lcnn_2d}.}
    \label{tab:arch_lcnn_4d}
    \begin{ruledtabular}
        \scriptsize
        \begin{tabular}{l | l | l }
            $W^{(2 \times 2)}$ & & \\
            \hline
            \hline
            & Small \hspace{8em} & Medium \hspace{7em} \\
             \hline
            & \Conv{2}{6}{2} \hspace{.3cm} & \Conv{3}{6}{4} \hspace{.3cm} \\
            & \Conv{2}{2}{2} & \Conv{3}{4}{4} \\
            & \GTr & \GTr \\
            & \Lin{4}{1} & \Lin{8}{1} \\
             \hline
            $N_\mathrm{param} \hspace{.3cm}$ & 1,801 & 8,305  \\
            \hline
            \hline
            $W^{(4 \times 4)}$ & & \\
            \hline
            \hline
            & Small & Medium \\
             \hline
            & \Conv{2}{6}{2} & \Conv{3}{6}{4} \\
            & \Conv{2}{2}{2} & \Conv{3}{4}{4} \\
            & \Conv{3}{2}{2} & \Conv{4}{4}{4} \\
            & \Conv{3}{2}{2} & \Conv{4}{4}{4} \\
            & \GTr & \GTr \\
            & \Lin{4}{1} & \Lin{8}{1} \\
             \hline
             $N_\mathrm{param}$ & 2,109 & 14,377  \\
             \hline
            \hline
            $q^\mathrm{plaq}$ & & \\
            \hline
            \hline
            & Small &  \\
             \hline
            & \Conv{2}{6}{4} &  \\
            & \GTr &  \\
            & \Lin{8}{1} &  \\
             \hline
             $N_\mathrm{param}$ & 3,181 &   \\
        \end{tabular}
    \end{ruledtabular}
\end{table}

Details about our \LGCNN{} architectures are provided in tables~\ref{tab:arch_lcnn_2d} and~\ref{tab:arch_lcnn_4d}, including the number of trainable parameters $N_\mathrm{param}$. To facilitate a fair comparison, the sizes of the \LGCNN{} architectures (in terms of trainable parameters) are chosen similar to those of the baseline networks. 

As stated in section~\ref{sec:implementation}, we combine gauge equivariant convolutions and bilinear layers into a single operation named \LCB{}. In our tables \Conv{K}{$N_\mathrm{in}$}{$N_\mathrm{out}$} denotes such a combined convolutional and bilinear layer with (quadratic) kernel size $K$, $N_\mathrm{in}$ input and $N_\mathrm{out}$ output channels. The \Trace{} operation takes the trace of all $\mathcal{W}$ channels at every lattice site. Given a configuration $(\mathcal{U}, \mathcal{W})$ with $N_\mathcal{W}$ channels for the $\mathcal{W}$ variables, the \Trace{} operation generates $2 \cdot N_\mathcal{W}$ real numbers at every lattice site. The doubling of the channel number is due to treating real and imaginary parts as separate channels. The outputs of the \Trace{} operation are then fed to single linear layers, which are applied at every lattice site individually. Therefore, the output of the \LGCNN{} (in contrast to our baseline models) is defined at every individual lattice site, which allows us to make predictions for e.g.~Wilson loops for the whole lattice configuration (i.e.~without the use of a lattice average or a global average pooling layer). Similar to our baseline networks, our \LGCNN{}s are translationally equivariant. Performing translations on the input field configuration leads to an appropriate shift of the prediction. For comparison to baseline predictions, a global lattice average is performed over the final output layer of the \LGCNN{} networks. In our studies we found that leaving out the lattice average for \LGCNN{}s led to much easier training in terms of convergence. \LGCNN{}s with global average pooling often did not converge at all. On the other hand, baselines trained without lattice average led to overall worse performance during training and testing. Test results for selected baseline architectures without lattice average are shown in table~\ref{tab:results_nogap}.

In the case of the smallest Wilson loop, $W^{(1 \times 1)}$, we have only used a single, trivial \LGCNN{} architecture consisting of a minimal \LCB{} and a \Trace{} operation. Since \LGCNN{} models are provided with both links and plaquettes ($\mathcal{U}$, $\mathcal{W}$), the calculation of the correct label $W^{(1 \times 1)}$ is trivial. For larger loops, the number of required \LCB{} operations grows with the area of the loop. With every application of an \LCB{} layer, the area of the loop is increased: starting from elementary \mbox{$1 \times 1$} and assuming a network with $n$ consecutive \LCB{} layers, the maximum possible loop area is $2^n$. Conversely, the number of required layers for an \mbox{$N \times N$} loop is therefore given by
\begin{align}
    n = \lceil \log_2 (N^2) \rceil,
\end{align}
where $\lceil \dots \rceil$ denotes the integer ceiling function.

\subsection{Training details}
Training of our \LGCNN{} models was performed using an \textit{NVidia RTX 2070 Super} (eight gigabytes) for two-dimensional models and lattices and an \textit{NVidia Titan V} (twelve gigabytes) for four-dimensional models and lattices. As in the  baseline study, each individual architecture shown in table~\ref{tab:arch_lcnn_2d} is trained ten times with randomly initialized parameters. Models for $W^{(1 \times 1)}$ and $W^{(1 \times 2)}$ are trained for a maximum of 20 epochs with a batch size of 50 and early stopping (patience value 5). The learning rate was set to $3 \cdot 10^{-3}$. Models for $W^{(2 \times 2)}$ and $W^{(4 \times 4)}$ are trained for a maximum of 100 epochs using a batch size of 50, early stopping (patience value 25) and a learning rate of $1 \cdot 10^{-3}$. In total, for our comparison study in 1+1D we train 100 individual \LGCNN{} models. Due to increased computational effort associated with 3+1D data, we train only five random instances of the architectures shown in table~\ref{tab:arch_lcnn_4d} using a batch size of 10. In total, this amounts to $25$ individual \LGCNN{} models. We use a learning rate of $3 \cdot 10^{-3}$ for $W^{(2 \times 2)}$ and $W^{(4 \times 4)}$ and a learning rate of $3 \cdot 10^{-4}$ for the topological charge $Q_P$. As discussed in section~\ref{sec:datasets}, we use a normalization factor for $Q_P$ to improve training. We use the \textit{AdamW} optimizer and zero weight decay for all of our models.

The current implementation of our  \LGCNN{} framework in \textit{PyTorch} typically requires larger computational resources than would be necessary for traditional CNNs. To give some impressions, the smallest \LGCNN{} for $W^{(1 \times 2)}$ in 1+1D (see table~\ref{tab:arch_lcnn_2d}) needs  $\approx 17$ seconds for a single epoch and slightly more than one gigabyte of GPU memory on a \textit{Titan V} GPU. The largest architecture for $W^{(4 \times 4)}$ loops in 1+1D takes $\approx 36$ seconds per epoch and close to three gigabytes of memory. Computational effort is drastically increased for 3+1D networks. Our largest $W^{(4 \times 4)}$ network in table~\ref{tab:arch_lcnn_4d} takes $\approx 13$ minutes for a single epoch and eight gigabytes of memory, or $\approx 11$ hours to train for $50$ epochs. Additional code optimizations should be able to improve the training efficiency of \LGCNN{}s further.

\subsection{Generation of Wilson loops}
As long as only the basic layers \Plaq{}, \LConv{}, \LBL{} and \Trace{} are involved, the set of locally transforming $\mathcal{W}$ objects generated by the \LGCNN{}s correspond to linear combinations of closed Wilson loops of growing length. To give a flavor of the complexity of the task these networks can in principle solve, we show in the following that the number of possible closed Wilson loops grows exponentially with the length of the loop. The \LGCNN{} networks systematically generate an increasing number of these loops with growing number of layers and channels. They share the advantageous properties of traditional CNNs regarding locality and weight sharing, but in a fully gauge-equivariant manner. The sketch of the proof in Fig.~2 of the Letter shows that with a sufficient number of layers in principle any contractible Wilson loop can be generated. We note that the proposed additional layers \LAct{} and \LExp{} can further extend the set of possible functions that the networks can express, beyond linear combinations of arbitrary Wilson loops. 

\begin{table}
    \caption{Number of possible closed loops on the lattice for a given loop length. The column labelled ``Closed random walk'' shows the number of different closed random walks of a given length. They may contain ``appendices'' where parts of the path may cancel trivially when evaluated on link configurations, because they go forward and immediately backward. ``Untraced'' shows the number of different closed loops of the given length which do not contain ``appendices'', but may contain a ``lasso'', a narrow line from the start/end point to the first loop-like structure.  ``Traced'' shows the number of different loops after a trace operation. \label{tab:number_of_loops}}
    \begin{ruledtabular}
        \scriptsize
        \begin{tabular}{l | r | r | r | r}
            Dim & Length & Closed random walk & Untraced & Traced \\
            \hline
            \hline
            2D &  2 &           4 &         0 &      0 \\
               &  4 &          36 &         8 &      2 \\
               &  6 &         400 &        40 &      4 \\
               &  8 &       4,900 &       312 &     28 \\
               & 10 &      63,504 &     2,240 &    152 \\
               & 12 &     853,776 &    17,280 &  1,010 \\
               & 14 &  11,778,624 &   134,568 &  6,772 \\
               & 16 & 165,636,900 & 1,071,000 & 47,646 \\
            
            \hline
            3D &  2 &              6 &           0 &          0 \\
               &  4 &             90 &          24 &          6 \\
               &  6 &          1,860 &         360 &         44 \\
               &  8 &         44,730 &       6,120 &        576 \\
               & 10 &      1,172,556 &     114,576 &      8,856 \\
               & 12 &     32,496,156 &   2,235,120 &    145,926 \\
               & 14 &    936,369,720 &  45,158,904 &  2,552,436 \\
               & 16 & 27,770,358,330 & 936,789,000 & 46,670,826 \\
            
            \hline
            4D &  2 &               8 &              0 &             0 \\
               &  4 &             168 &             48 &            12 \\
               &  6 &           5,120 &          1,200 &           152 \\
               &  8 &         190,120 &         36,432 &         3,624 \\
               & 10 &       7,939,008 &      1,202,880 &        97,680 \\
               & 12 &     357,713,664 &     42,396,480 &     2,912,844 \\
               & 14 &  16,993,726,464 &  1,564,370,928 &    93,039,192 \\
               & 16 & 839,358,285,480 & 59,773,380,240 & 3,132,835,092 \\
        \end{tabular}
    \end{ruledtabular}
\end{table}

Table~\ref{tab:number_of_loops} shows the number of different possible Wilson loops for a given length, that is the number of links that have to be multiplied to form a loop. For this analysis, we assume an infinitely large lattice. The reason to count the number of possible loops according to loop length and not to area is that for a given area, loops can have an unbound length, for example two plaquettes connected by a pair of long Wilson lines that lie on top of each other and therefore enclose a vanishing surface. On the other hand, for a given length of the loop, the number of possible loops on a lattice is finite, but grows exponentially with the loop length.

One can provide formulae for the number of open and closed simple random walks of given length $L$. The number of simple random walks in $D$ dimensions is $(2D)^L$, because at each of the $L$ steps there are $2D$ possible directions (along the positive or negative axis along each of the $D$ dimensions). Most of these paths will not return to the starting point though. The number of closed random walks, that is those that return to the origin of the walk, is given by
\begin{align}
    D=1: \quad & \frac{L!}{\left(\left(\frac{L}{2}\right)!\right)^2} = {L \choose \frac{L}{2}}, \\
    D=2: \quad & \sum_{i=0}^{\frac{L}{2}} \frac{L!}{(i!)^2 \left(\left(\frac{L}{2} - i \right)!\right)^2}  = {L \choose \frac{L}{2}}^2, \\
    D=3: \quad & \sum_{i=0}^{\frac{L}{2}} \sum_{j=0}^{\frac{L}{2} - i} \frac{L!}{(i!)^2 (j!)^2 \left(\left( \frac{L}{2} - i - j\right)!\right)^2}, \\
    D=4: \quad & \sum_{i=0}^{\frac{L}{2}} \sum_{j=0}^{\frac{L}{2} - i} \sum_{k=0}^{\frac{L}{2}\mspace{-2mu} -\mspace{-2mu} i\mspace{-2mu} -\mspace{-2mu} j} \frac{L!}{(i!)^2 (j!)^2 (k!)^2 \left(\left( \frac{L}{2}\! -\! i\! -\! j\! -\! k\right)!\right)^2},
\end{align}
for even $L$. For odd $L$ the random path cannot return to the origin. These formulae can be obtained by realizing that for closed loops the number of links pointing in one direction and the number of links pointing in the opposite direction must be equal for each dimension separately. The sums produce all possible combinations of such link pairs while the fractional expression counts the number of possibilities to choose the given number of links for positive and negative directions from all links. In $D=1$ and $D=2$ dimensions, simpler expressions in terms of binomial coefficients are possible. We note that there is a famous theorem in this context which states that the probability of an infinitely long simple random walk to return to the origin at some point equals 1 up to $D\leq 2$ (``recurrence''), but is less than 1 for $D>2$ (``transience''). This has first been investigated by P\'olya in 1921~\cite{Polya:1921}.

Many of these closed random walks may contain ``appendices'' where parts of the path may cancel trivially if they are evaluated as Wilson loops on a link configuration. This is because multiplying a link with the same link backwards just gives $U U^\dagger = \mathbb{1}$. After removing these appendices from the diagrams, we will refer to the set of remaining diagrams as ``untraced'' Wilson loops. It is difficult to provide a closed formula for the number of untraced Wilson loops due to subtle combinatorics. It is straightforward though to create all possible loops of given length programmatically and count the resulting number of loops. Tracing the diagrams can create new appendices to the origin if the untraced diagram contained a ``lasso''-like structure. Untraced diagrams with different starting points within a closed loop may correspond to the same traced diagram, so in order to count the number of traced diagrams properly, possible duplicates have to be removed first. The results are shown in Table~\ref{tab:number_of_loops}. 

\begin{figure}
    \centering
    \includegraphics[width=0.47\textwidth]{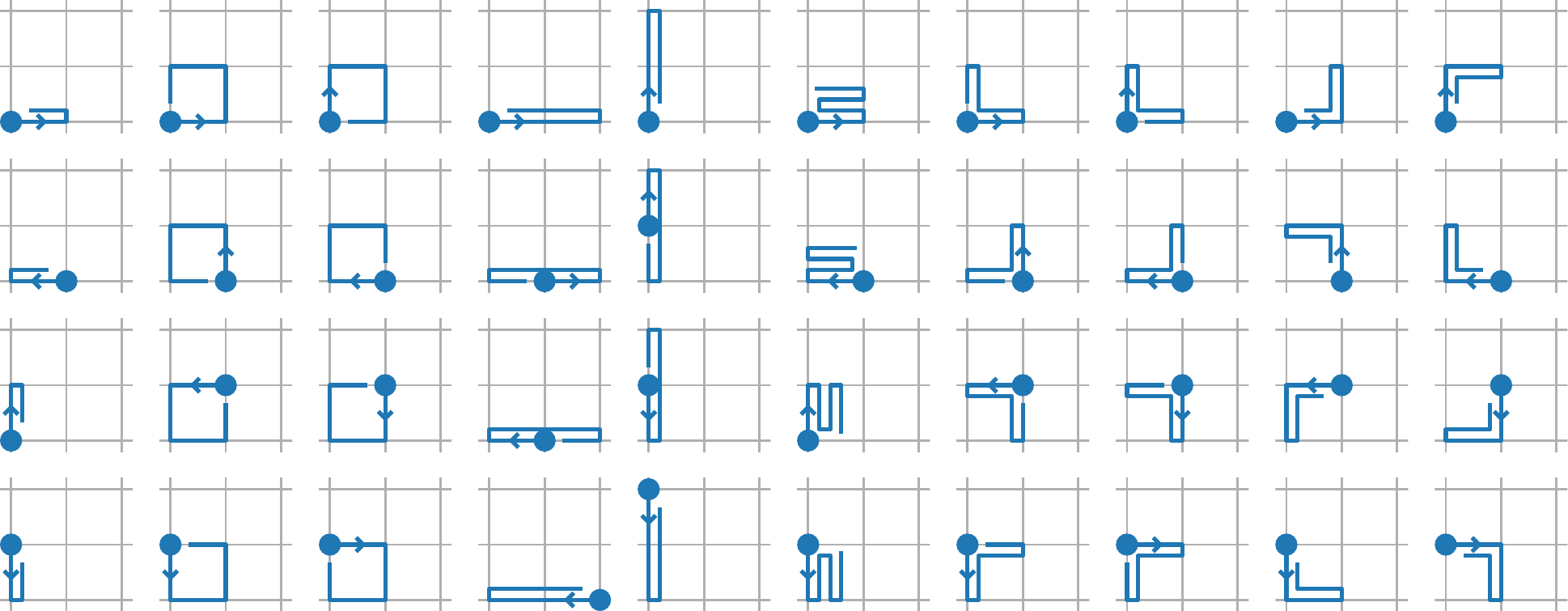}
    \caption{All closed random walk diagrams in two dimensions up to length 4. The first column shows all four random walk diagrams of length 2. The remaining columns show the 36 diagrams of length 4. The blue dot marks the starting and end point of the random walk. The diagrams in the sixth column indicate going backward and forward twice along the same link. When evaluated on links, many path segments cancel. We refer to the remaining set of unique diagrams as ``untraced'' closed Wilson loops (which for this set of diagrams contains the 8 untraced plaquette diagrams in the second and third column.)}
    \label{fig:loops-random}
\end{figure}

\begin{figure}
    \centering
    \includegraphics[width=0.47\textwidth]{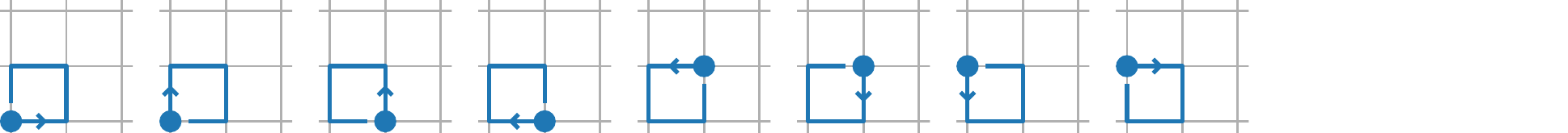}
    \vspace{.18cm}
    \includegraphics[width=0.47\textwidth]{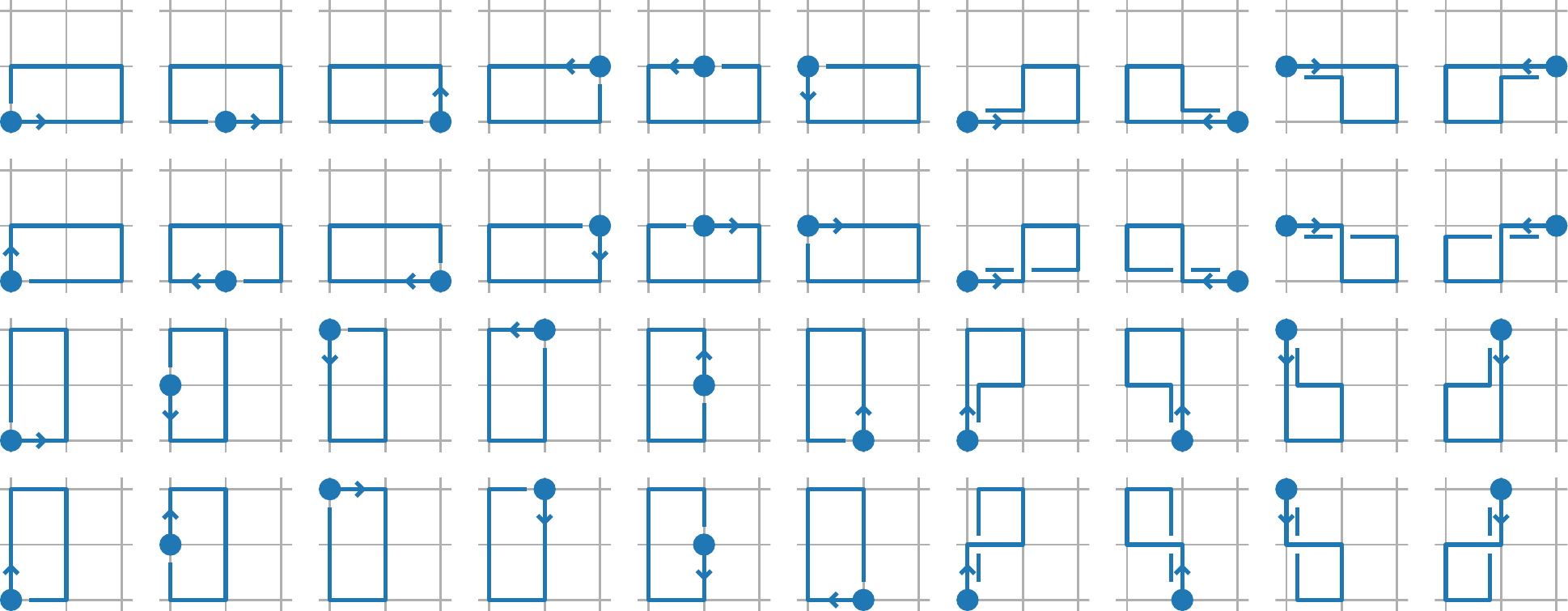}
    \caption{All untraced diagrams in two dimensions up to length 6. The first row shows all diagrams of length 4. The second to fifth row show the 40 diagrams of length 6. The blue dot marks the starting point where the closed but untraced Wilson loop transforms.}
    \label{fig:loops-untraced}
\end{figure}

\begin{figure}
    \centering
    \includegraphics[width=0.47\textwidth]{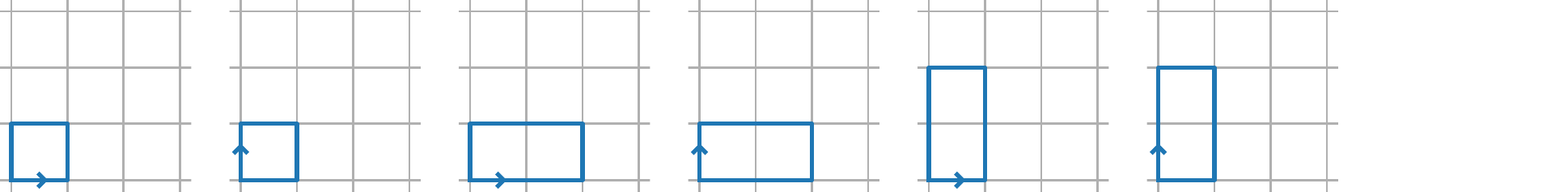}
    \vspace{.18cm}
    \includegraphics[width=0.47\textwidth]{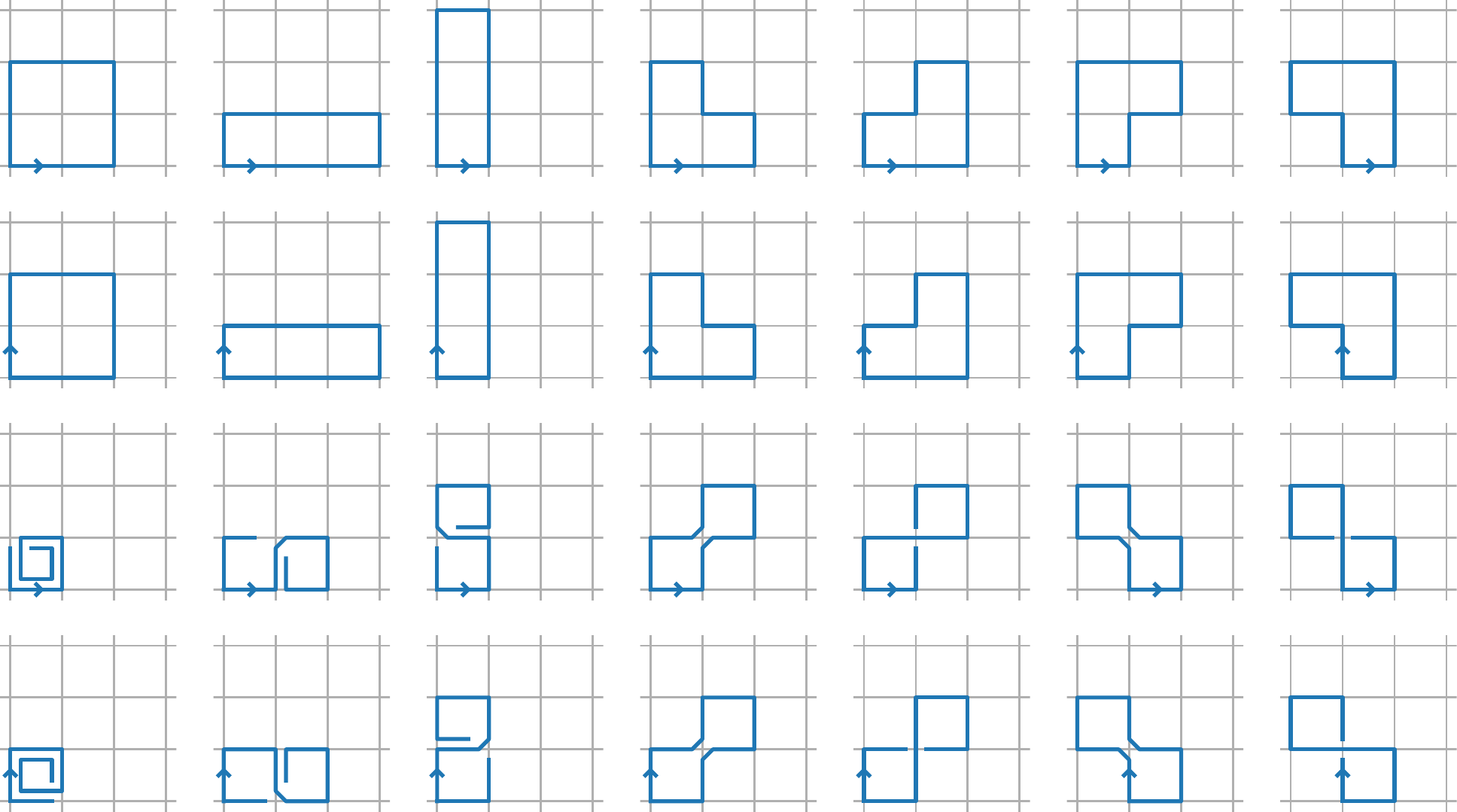}
    \caption{All traced diagrams in two dimensions up to length 8. The first row shows all diagrams of length 4 and 6. The second to fifth row show the 28 diagrams of length 8. Each of the diagrams has a corresponding diagram with opposite orientation within this set which is its Hermitian conjugate. }
    \label{fig:loops-traced}
\end{figure}

Figures~\ref{fig:loops-random}, \ref{fig:loops-untraced}, and~\ref{fig:loops-traced} illustrate the first few closed random loops, untraced loops, and traced loops, respectively. After removing the appendices of all diagrams of Fig.~\ref{fig:loops-random}, only the diagrams in the first row of Fig.~\ref{fig:loops-untraced} remain; after tracing all diagrams of Fig.~\ref{fig:loops-untraced}, only the diagrams in the first row of Fig.~\ref{fig:loops-traced} remain.

\begin{table}
    \caption{Number of untraced loops covered by the \LGCNN{} architectures considered, for small (S), medium (M) and large (L) architectures for various tasks in 1+1D as given in table~\ref{tab:arch_lcnn_2d}.}
    \label{tab:untraced_loops_for_architectures}
    \begin{ruledtabular}
        \scriptsize
        \begin{tabular}{l | r | r | r | r | r | r | r | r }
            Length & Max & $W^{(1\times1)}$ & \multicolumn{3}{l|}{$W^{(1 \times 2)}$} & \multicolumn{3}{l}{$W^{(2 \times 2)}$}  \\
               &            & S & S & M &  L &   S &     M &     L \\
            \hline
            \hline
             0 &           1 & 1 & 1 & 1 &  1 &   1 &     1 &      1 \\
             2 &           0 & 0 & 0 & 0 &  0 &   0 &     0 &      0 \\
             4 &           8 & 2 & 2 & 2 &  2 &   2 &     2 &      2 \\
             6 &          40 &   & 6 & 6 &  6 &   8 &     8 &      8 \\
             8 &         312 &   & 2 & 6 &  6 &  34 &    34 &     34 \\
            10 &       2,240 &   & 4 & 8 & 12 &  54 &   104 &    106 \\
            12 &      17,280 &   &   & 4 &  8 & 112 &   226 &    268 \\
            14 &     134,568 &   &   &   &  4 & 166 &   472 &    618 \\
            16 &   1,071,000 &   &   &   &    & 182 &   760 &  1,182 \\
            18 &   8,627,872 &   &   &   &    & 196 & 1,114 &  2,040 \\
            20 &  70,302,888 &   &   &   &    &  96 & 1,282 &  2,892 \\
            22 & 577,920,200 &   &   &   &    &  64 & 1,192 &  3,528 \\
            $\geq$24 &       &   &   &   &    &   0 & 1,248 & 10,228 \\
            \hline
            Total &          & 3 &15 &27 & 39 & 915 & 6,443 & 20,907 \\
            \hline
            \hline
            Max.Len &        & 4 &10 &12 & 14 &  22 &    28 &     34     
        \end{tabular}
    \end{ruledtabular}
\end{table}

\begin{table}
    \caption{Number of traced loops covered by the \LGCNN{} architectures considered, for small (S), medium (M) and large (L) architectures for various tasks in 1+1D as given in table~\ref{tab:arch_lcnn_2d}.}
    \label{tab:traced_loops_for_architectures}
    \begin{ruledtabular}
        \scriptsize
        \begin{tabular}{l | r | r | r | r | r | r | r | r }
            Length & Max & $W^{(1\times1)}$ & \multicolumn{3}{l|}{$W^{(1 \times 2)}$} & \multicolumn{3}{l}{$W^{(2 \times 2)}$}  \\
                   &     & S & S & M & L & S &    M &     L \\
            \hline
            \hline
             0 &          1 & 1 & 1 & 1 & 1 &   1 &     1 &     1 \\
             2 &          0 & 0 & 0 & 0 & 0 &   0 &     0 &     0 \\
             4 &          2 & 2 & 2 & 2 & 2 &   2 &     2 &     2 \\
             6 &          4 &   & 4 & 4 & 4 &   4 &     4 &     4 \\
             8 &         28 &   & 4 & 4 & 4 &  22 &    22 &    22 \\
            10 &        152 &   &   & 8 & 8 &  48 &    76 &    76 \\
            12 &      1,010 &   &   &   & 8 &  92 &   204 &   220 \\
            14 &      6,772 &   &   &   &   & 120 &   412 &   532 \\
            16 &     47,646 &   &   &   &   & 100 &   712 &  1,080 \\
            18 &    343,168 &   &   &   &   & 136 &   928 &  1,896 \\
            20 &  2,529,890 &   &   &   &   &  32 & 1,056 &  2,620 \\
            22 & 18,982,172 &   &   &   &   &  64 &   768 &  3,152 \\
            $\geq$24 &      &   &   &   &   &     &   800 &  7,210 \\
            \hline
            Total &        & 3 & 11& 19& 27& 621 & 4,985 & 16,725 \\
            \hline
            \hline
            Max.Len &      & 4 & 8 & 10& 12&  22 &    28 &     34
        \end{tabular}
    \end{ruledtabular}
\end{table}

Tables~\ref{tab:untraced_loops_for_architectures} and~\ref{tab:traced_loops_for_architectures} show how many loops can in principle be generated by the \LGCNN{} architectures used for our tasks as specified in table~\ref{tab:arch_lcnn_2d}. The tasks we chose have the following loop lengths: $W^{(1 \times 1)}$ contains 4 links, $W^{(1 \times 2)}$ 6 links, $W^{(2 \times 2)}$ 8 links, and $W^{(4 \times 4)}$ 16 links. The topological charge task $q^\mathrm{plaq}_x$ requires linear combinations of traced Wilson loops of length 8. The \LConv{} and \LBL{} layers (or the combined \LCB{} operation) form an increasing number of untraced Wilson loops at each layer. The Trace layer forms traced Wilson loops in the end. Since our sample tasks only involved loops in the first quadrant, we restricted ourselves to positive shifts in the \LConv{} layers. This means that from the 8 possible untraced loops of length 4 in the first row of Fig.~\ref{fig:loops-untraced} only the first two plaquettes are covered. From the 40 possible diagrams of length 6, only 8 diagrams are covered by the networks with two convolutions, which are those diagrams with the starting point in the lower left corner. With only one convolution, the second and the fourth diagram in the first column of Fig.~\ref{fig:loops-untraced} are not generated such that only 6 diagrams are covered. While this choice of using only positive shifts may be limiting for untraced loops, after taking the trace, the starting point of the loops does not matter, and all traced loops up to length 6 are indeed covered as can be seen in table~\ref{tab:traced_loops_for_architectures}. Figure~\ref{fig:loops-traced} shows the traced diagrams up to length 8. The \mbox{$1 \times 1$}, \mbox{$1 \times 2$} and \mbox{$2 \times 2$} Wilson loops are covered by the \LGCNN{} architectures for the respective tasks. Six of these diagrams are not captured by the chosen architectures (the 4 diagrams in the last column and the 2 lowest diagrams in the penultimate column of Fig.~\ref{fig:loops-traced}, which are the 6 diagrams that do not touch the lower left corner of the gray background grid), because we restricted our networks to positive shifts in the convolution. They would be covered as well, if negative shifts were also included or if an additional \LCB{} layer was used, as shown in the following paragraph.

With our existing approach, we have so far not been able to perform the corresponding loop analysis for the networks for the $W^{(4 \times 4)}$ task in 1+1D or 3+1D. This is because the number of possible loops is roughly squared with each additional bilinear layer, with even larger growth for larger kernel sizes in the convolution. It is still possible to study networks of three \LCB{} layers which lie in size between the networks for the $W^{(2 \times 2)}$ and $W^{(4 \times 4)}$ tasks. In 1+1D, the small $W^{(2 \times 2)}$ architecture with one additional L-CB(2,2,2) layer covers a total of 4,967,355 untraced or 3,780,443 traced loops. All 28 traced loops of length 8 are covered (which was not the case yet with only 2 \LCB{} layers), as well as 128 (454; 1592) traced loops of length 10 (12; 14), with a maximum traced (untraced) loop length of 44 (46). It is future work to study also larger networks with more than three \LCB{} layers in more detail, including the architectures for $W^{(4 \times 4)}$ tasks. Following the trend, the total number of loops for four \LCB{} layers will be more than the square of the corresponding numbers for three layers and likely exceed $10^{13}$ diagrams.

Finally, we point out again that \LGCNN{}s do not literally generate all these diagrams, but all of these possible diagrams can in principle be represented by adjusting the weights of the network layers properly, which is done by backpropagation and stochastic gradient descent. An \LGCNN{} network processes a set of $\mathcal W$ objects which correspond to linear combinations of Wilson loops as long as only \LConv{} and \LBL{} are involved. This is why the number of trainable parameters given in table~\ref{tab:arch_lcnn_2d} can be smaller than the number of Wilson loops they can potentially reproduce. The structure of convolutional and bilinear operations, or the structure of the combined layer as can be seen for the trainable parameters $\alpha$ in eq.~\eqref{eq:lcb}, determines which combinations of loops can be formed. Only for a large enough number of channels it is possible to reproduce each diagram separately, and with a sufficient number of layers, all possible loops for a given length can be covered in principle.

\section{Test results} \label{sec:results}
\begin{table}[ht]
    \caption{Test results for \LGCNN{} and baseline CNN architectures (denoted as ``Base'') on the $W^{(1\times 1)}$ regression task in 1+1D. The \LGCNN{}s are provided with links and single-orientation plaquettes as input $(\mathcal{U}, \mathcal{W})$. The baseline models are only provided with gauge links $\mathcal{U}$. This table shows the median mean squared error of the model ensemble for each architecture type and each lattice size. Lowest values are highlighted in boldface. In the case of baseline CNNs, we also add the type of activation function used. Architecture details are provided in tables~\ref{tab:arch_base_w1x2} and~\ref{tab:arch_lcnn_2d}. We also report the variance of labels (observables) in the test data set.}
    \label{tab:results_1x1_u}
    \scriptsize
    \begin{ruledtabular}
        \begin{tabular}{l | l | l | l | l}
             & $8 \cdot 8$ & $16 \cdot 16$ & $32 \cdot 32$ & $64 \cdot 64$ \\
            \hline
            Variance & \nnum{5.97e-02} & \nnum{5.69e-02} & \nnum{5.64e-02} & \nnum{5.61e-02} \\
            \hline
            \LGCNN{}& \bnum{2.19e-08} & \bnum{2.19e-08} & \bnum{2.19e-08} & \bnum{2.19e-08}  \\ 
            \hline 
            Base S1 (tanh)& \nnum{5.40e-02} & \nnum{4.74e-02} & \nnum{4.51e-02} & \nnum{4.48e-02}  \\ 
            Base S1 (sigm)& \nnum{5.89e-02} & \nnum{5.48e-02} & \nnum{5.37e-02} & \nnum{5.33e-02}  \\ 
            Base S1 (leaky)& \nnum{4.88e-02} & \nnum{4.00e-02} & \nnum{3.74e-02} & \nnum{3.69e-02}  \\ 
            Base S1 (relu)& \nnum{5.97e-02} & \nnum{5.69e-02} & \nnum{5.64e-02} & \nnum{5.61e-02}  \\ 
            \hline 
            Base S2 (tanh)& \nnum{5.98e-02} & \nnum{5.68e-02} & \nnum{5.63e-02} & \nnum{5.61e-02}  \\ 
            Base S2 (sigm)& \nnum{5.97e-02} & \nnum{5.69e-02} & \nnum{5.63e-02} & \nnum{5.61e-02}  \\ 
            Base S2 (leaky)& \nnum{5.06e-02} & \nnum{4.18e-02} & \nnum{3.99e-02} & \nnum{3.91e-02}  \\ 
            Base S2 (relu)& \nnum{5.97e-02} & \nnum{5.69e-02} & \nnum{5.64e-02} & \nnum{5.61e-02}  \\ 
            \hline 
            Base S3 (tanh)& \nnum{5.82e-02} & \nnum{5.09e-02} & \nnum{4.96e-02} & \nnum{4.93e-02}  \\ 
            Base S3 (sigm)& \nnum{5.97e-02} & \nnum{5.69e-02} & \nnum{5.64e-02} & \nnum{5.61e-02}  \\ 
            Base S3 (leaky)& \nnum{5.97e-02} & \nnum{5.69e-02} & \nnum{5.64e-02} & \nnum{5.61e-02}  \\ 
            Base S3 (relu)& \nnum{5.95e-02} & \nnum{5.64e-02} & \nnum{5.59e-02} & \nnum{5.57e-02}  \\ 
            \hline 
            Base M1 (tanh)& \nnum{5.97e-02} & \nnum{5.69e-02} & \nnum{5.64e-02} & \nnum{5.61e-02}  \\ 
            Base M1 (sigm)& \nnum{5.97e-02} & \nnum{5.69e-02} & \nnum{5.64e-02} & \nnum{5.61e-02}  \\ 
            Base M1 (leaky)& \nnum{5.11e-02} & \nnum{4.52e-02} & \nnum{4.34e-02} & \nnum{4.29e-02}  \\ 
            Base M1 (relu)& \nnum{5.97e-02} & \nnum{5.69e-02} & \nnum{5.64e-02} & \nnum{5.61e-02}  \\ 
            \hline 
            Base M2 (tanh)& \nnum{5.99e-02} & \nnum{5.69e-02} & \nnum{5.64e-02} & \nnum{5.62e-02}  \\ 
            Base M2 (sigm)& \nnum{5.92e-02} & \nnum{5.27e-02} & \nnum{5.12e-02} & \nnum{5.07e-02}  \\ 
            Base M2 (leaky)& \nnum{2.55e-02} & \nnum{1.28e-02} & \nnum{1.02e-02} & \nnum{9.55e-03}  \\ 
            Base M2 (relu)& \nnum{3.72e-02} & \nnum{2.47e-02} & \nnum{2.18e-02} & \nnum{2.12e-02}  \\ 
            \hline 
            Base M3 (tanh)& \nnum{5.98e-02} & \nnum{5.69e-02} & \nnum{5.64e-02} & \nnum{5.61e-02}  \\ 
            Base M3 (sigm)& \nnum{5.97e-02} & \nnum{5.69e-02} & \nnum{5.64e-02} & \nnum{5.61e-02}  \\ 
            Base M3 (leaky)& \nnum{5.97e-02} & \nnum{5.69e-02} & \nnum{5.63e-02} & \nnum{5.61e-02}  \\ 
            Base M3 (relu)& \nnum{5.97e-02} & \nnum{5.69e-02} & \nnum{5.64e-02} & \nnum{5.61e-02}  \\ 
            \hline 
            Base L1 (tanh)& \nnum{5.99e-02} & \nnum{5.70e-02} & \nnum{5.64e-02} & \nnum{5.62e-02}  \\ 
            Base L1 (sigm)& \nnum{5.97e-02} & \nnum{5.69e-02} & \nnum{5.64e-02} & \nnum{5.61e-02}  \\ 
            Base L1 (leaky)& \nnum{5.97e-02} & \nnum{5.69e-02} & \nnum{5.64e-02} & \nnum{5.61e-02}  \\ 
            Base L1 (relu)& \nnum{5.97e-02} & \nnum{5.69e-02} & \nnum{5.64e-02} & \nnum{5.61e-02}  \\ 
            \hline 
            Base L2 (tanh)& \nnum{6.00e-02} & \nnum{5.70e-02} & \nnum{5.64e-02} & \nnum{5.61e-02}  \\ 
            Base L2 (sigm)& \nnum{5.98e-02} & \nnum{5.69e-02} & \nnum{5.64e-02} & \nnum{5.61e-02}  \\ 
            Base L2 (leaky)& \nnum{5.98e-02} & \nnum{5.69e-02} & \nnum{5.63e-02} & \nnum{5.61e-02}  \\ 
            Base L2 (relu)& \nnum{5.97e-02} & \nnum{5.69e-02} & \nnum{5.64e-02} & \nnum{5.61e-02}  \\ 
            \hline 
            Base L3 (tanh)& \nnum{5.98e-02} & \nnum{5.69e-02} & \nnum{5.64e-02} & \nnum{5.62e-02}  \\ 
            Base L3 (sigm)& \nnum{5.97e-02} & \nnum{5.69e-02} & \nnum{5.64e-02} & \nnum{5.61e-02}  \\ 
            Base L3 (leaky)& \nnum{5.97e-02} & \nnum{5.69e-02} & \nnum{5.63e-02} & \nnum{5.61e-02}  \\ 
            Base L3 (relu)& \nnum{5.97e-02} & \nnum{5.69e-02} & \nnum{5.64e-02} & \nnum{5.61e-02}  \\ 
            \hline 
            Base W1 (tanh)& \nnum{6.19e-02} & \nnum{5.75e-02} & \nnum{5.66e-02} & \nnum{5.62e-02}  \\ 
            Base W1 (sigm)& \nnum{5.93e-02} & \nnum{5.66e-02} & \nnum{5.61e-02} & \nnum{5.59e-02}  \\ 
            Base W1 (leaky)& \nnum{1.88e-02} & \nnum{8.05e-03} & \nnum{5.79e-03} & \nnum{5.11e-03}  \\ 
            Base W1 (relu)& \nnum{2.08e-02} & \nnum{9.02e-03} & \nnum{6.53e-03} & \nnum{5.77e-03}  \\ 
            \hline 
            Base W2 (tanh)& \nnum{6.01e-02} & \nnum{5.70e-02} & \nnum{5.62e-02} & \nnum{5.60e-02}  \\ 
            Base W2 (sigm)& \nnum{5.97e-02} & \nnum{5.69e-02} & \nnum{5.64e-02} & \nnum{5.61e-02}  \\ 
            Base W2 (leaky)& \nnum{2.11e-02} & \nnum{1.06e-02} & \nnum{8.23e-03} & \nnum{7.62e-03}  \\ 
            Base W2 (relu)& \nnum{5.97e-02} & \nnum{5.69e-02} & \nnum{5.64e-02} & \nnum{5.61e-02}  \\ 
            \hline 
            Base W3 (tanh)& \nnum{6.03e-02} & \nnum{5.75e-02} & \nnum{5.66e-02} & \nnum{5.63e-02}  \\ 
            Base W3 (sigm)& \nnum{5.97e-02} & \nnum{5.69e-02} & \nnum{5.64e-02} & \nnum{5.61e-02}  \\ 
            Base W3 (leaky)& \nnum{2.15e-02} & \nnum{1.17e-02} & \nnum{9.80e-03} & \nnum{9.14e-03}  \\ 
            Base W3 (relu)& \nnum{5.96e-02} & \nnum{5.63e-02} & \nnum{5.57e-02} & \nnum{5.55e-02}  \\ 
        \end{tabular}
    \end{ruledtabular}
\end{table}

\begin{table}[ht]
    \caption{Test results for \LGCNN{} and baseline CNN architectures (denoted as ``Base'') on the $W^{(1\times 1)}$ regression task in 1+1D. The \LGCNN{} and the baseline models are provided with links and single-orientation plaquettes as input $(\mathcal{U}, \mathcal{W})$. Architecture details are provided in tables~\ref{tab:arch_base_w1x2} and~\ref{tab:arch_lcnn_2d}.}
    \label{tab:results_1x1_uw}
    \scriptsize
    \begin{ruledtabular}
        \begin{tabular}{l | l | l | l | l}
             & $8 \cdot 8$ & $16 \cdot 16$ & $32 \cdot 32$  & $64 \cdot 64$   \\
            \hline
            Variance & \nnum{5.97e-02} & \nnum{5.69e-02} & \nnum{5.64e-02} & \nnum{5.61e-02} \\
            \hline
            \LGCNN{}& \bnum{2.19e-08} & \nnum{2.19e-08} & \nnum{2.19e-08} & \nnum{2.19e-08}  \\ 
            \hline 
            Base S1 (tanh)& \nnum{2.00e-05} & \nnum{5.64e-06} & \nnum{2.33e-06} & \nnum{1.65e-06}  \\ 
            Base S1 (sigm)& \nnum{6.88e-06} & \nnum{2.12e-06} & \nnum{9.56e-07} & \nnum{7.42e-07}  \\ 
            Base S1 (leaky)& \nnum{5.87e-08} & \nnum{1.53e-08} & \bnum{6.95e-09} & \nnum{4.61e-09}  \\ 
            Base S1 (relu)& \nnum{3.67e-08} & \bnum{1.33e-08} & \nnum{6.99e-09} & \nnum{5.63e-09}  \\ 
            \hline 
            Base S2 (tanh)& \nnum{2.84e-06} & \nnum{8.99e-07} & \nnum{4.60e-07} & \nnum{3.86e-07}  \\ 
            Base S2 (sigm)& \nnum{6.32e-06} & \nnum{2.42e-06} & \nnum{1.39e-06} & \nnum{9.32e-07}  \\ 
            Base S2 (leaky)& \nnum{9.65e-08} & \nnum{2.93e-08} & \nnum{7.33e-09} & \bnum{2.98e-09}  \\ 
            Base S2 (relu)& \nnum{1.11e-06} & \nnum{3.40e-07} & \nnum{9.14e-08} & \nnum{5.86e-08}  \\ 
            \hline 
            Base S3 (tanh)& \nnum{1.42e-06} & \nnum{5.23e-07} & \nnum{2.74e-07} & \nnum{2.12e-07}  \\ 
            Base S3 (sigm)& \nnum{2.22e-06} & \nnum{6.46e-07} & \nnum{2.63e-07} & \nnum{1.74e-07}  \\ 
            Base S3 (leaky)& \nnum{1.18e-07} & \nnum{4.76e-08} & \nnum{1.47e-08} & \nnum{1.10e-08}  \\ 
            Base S3 (relu)& \nnum{8.47e-07} & \nnum{3.79e-07} & \nnum{1.40e-07} & \nnum{1.13e-07}  \\ 
            \hline 
            Base M1 (tanh)& \nnum{1.22e-05} & \nnum{3.82e-06} & \nnum{1.34e-06} & \nnum{1.01e-06}  \\ 
            Base M1 (sigm)& \nnum{2.52e-05} & \nnum{7.47e-06} & \nnum{2.54e-06} & \nnum{1.72e-06}  \\ 
            Base M1 (leaky)& \nnum{4.22e-07} & \nnum{1.19e-07} & \nnum{5.73e-08} & \nnum{3.92e-08}  \\ 
            Base M1 (relu)& \nnum{3.92e-06} & \nnum{1.04e-06} & \nnum{2.51e-07} & \nnum{8.72e-08}  \\ 
            \hline 
            Base M2 (tanh)& \nnum{7.99e-06} & \nnum{2.69e-06} & \nnum{1.66e-06} & \nnum{1.31e-06}  \\ 
            Base M2 (sigm)& \nnum{4.61e-06} & \nnum{1.63e-06} & \nnum{7.87e-07} & \nnum{5.67e-07}  \\ 
            Base M2 (leaky)& \nnum{1.40e-07} & \nnum{4.20e-08} & \nnum{1.50e-08} & \nnum{9.45e-09}  \\ 
            Base M2 (relu)& \nnum{1.06e-06} & \nnum{3.39e-07} & \nnum{1.21e-07} & \nnum{5.40e-08}  \\ 
            \hline 
            Base M3 (tanh)& \nnum{1.43e-05} & \nnum{5.20e-06} & \nnum{2.48e-06} & \nnum{1.94e-06}  \\ 
            Base M3 (sigm)& \nnum{3.96e-06} & \nnum{1.53e-06} & \nnum{9.27e-07} & \nnum{7.89e-07}  \\ 
            Base M3 (leaky)& \nnum{1.55e-07} & \nnum{4.31e-08} & \nnum{1.73e-08} & \nnum{1.08e-08}  \\ 
            Base M3 (relu)& \nnum{1.39e-07} & \nnum{4.48e-08} & \nnum{2.28e-08} & \nnum{1.70e-08}  \\ 
            \hline 
            Base L1 (tanh)& \nnum{1.58e-05} & \nnum{5.13e-06} & \nnum{1.72e-06} & \nnum{1.24e-06}  \\ 
            Base L1 (sigm)& \nnum{2.58e-05} & \nnum{6.73e-06} & \nnum{3.90e-06} & \nnum{3.68e-06}  \\ 
            Base L1 (leaky)& \nnum{1.10e-05} & \nnum{2.90e-06} & \nnum{1.07e-06} & \nnum{6.56e-07}  \\ 
            Base L1 (relu)& \nnum{6.56e-06} & \nnum{2.14e-06} & \nnum{7.98e-07} & \nnum{4.87e-07}  \\ 
            \hline 
            Base L2 (tanh)& \nnum{3.35e-05} & \nnum{1.03e-05} & \nnum{3.58e-06} & \nnum{1.93e-06}  \\ 
            Base L2 (sigm)& \nnum{3.00e-02} & \nnum{2.85e-02} & \nnum{2.82e-02} & \nnum{2.81e-02}  \\ 
            Base L2 (leaky)& \nnum{1.56e-05} & \nnum{4.44e-06} & \nnum{1.37e-06} & \nnum{5.71e-07}  \\ 
            Base L2 (relu)& \nnum{1.34e-05} & \nnum{1.06e-05} & \nnum{6.74e-06} & \nnum{6.17e-06}  \\ 
            \hline 
            Base L3 (tanh)& \nnum{3.89e-05} & \nnum{1.15e-05} & \nnum{5.60e-06} & \nnum{4.17e-06}  \\ 
            Base L3 (sigm)& \nnum{2.66e-05} & \nnum{8.77e-06} & \nnum{4.20e-06} & \nnum{3.17e-06}  \\ 
            Base L3 (leaky)& \nnum{2.09e-06} & \nnum{6.01e-07} & \nnum{2.40e-07} & \nnum{1.34e-07}  \\ 
            Base L3 (relu)& \nnum{9.41e-06} & \nnum{2.39e-06} & \nnum{8.73e-07} & \nnum{3.85e-07}  \\ 
            \hline 
            Base W1 (tanh)& \nnum{1.21e-06} & \nnum{3.60e-07} & \nnum{1.40e-07} & \nnum{9.01e-08}  \\ 
            Base W1 (sigm)& \nnum{3.78e-06} & \nnum{1.08e-06} & \nnum{3.54e-07} & \nnum{1.49e-07}  \\ 
            Base W1 (leaky)& \nnum{2.40e-06} & \nnum{6.56e-07} & \nnum{2.16e-07} & \nnum{9.28e-08}  \\ 
            Base W1 (relu)& \nnum{2.41e-06} & \nnum{6.74e-07} & \nnum{2.42e-07} & \nnum{1.13e-07}  \\ 
            \hline 
            Base W2 (tanh)& \nnum{2.43e-04} & \nnum{6.42e-05} & \nnum{1.76e-05} & \nnum{7.67e-06}  \\ 
            Base W2 (sigm)& \nnum{5.97e-02} & \nnum{5.69e-02} & \nnum{5.64e-02} & \nnum{5.61e-02}  \\ 
            Base W2 (leaky)& \nnum{1.58e-05} & \nnum{5.03e-06} & \nnum{1.73e-06} & \nnum{9.48e-07}  \\ 
            Base W2 (relu)& \nnum{5.97e-02} & \nnum{5.69e-02} & \nnum{5.64e-02} & \nnum{5.61e-02}  \\ 
            \hline 
            Base W3 (tanh)& \nnum{4.50e-05} & \nnum{1.56e-05} & \nnum{8.74e-06} & \nnum{7.21e-06}  \\ 
            Base W3 (sigm)& \nnum{5.97e-02} & \nnum{5.69e-02} & \nnum{5.64e-02} & \nnum{5.61e-02}  \\ 
            Base W3 (leaky)& \nnum{1.15e-05} & \nnum{3.16e-06} & \nnum{1.28e-06} & \nnum{7.00e-07}  \\ 
            Base W3 (relu)& \nnum{2.04e-06} & \nnum{5.43e-07} & \nnum{2.07e-07} & \nnum{1.14e-07}  \\ 
        \end{tabular}
    \end{ruledtabular}
\end{table}

\begin{table}[ht]
    \caption{Test results for \LGCNN{} and baseline CNN architectures (denoted as ``Base'') on the $W^{(1\times 1)}$ regression task in 1+1D. \LGCNN{} models are provided with links and single-orientation plaquettes as input $(\mathcal{U}, \mathcal{W})$. Baseline models are provided with links and plaquettes of both orientations $(\mathcal{U}, \mathcal{W}, \mathcal{W}^\dg)$. Architecture details are provided in tables~\ref{tab:arch_base_w1x2} and~\ref{tab:arch_lcnn_2d}. The asterisk ($*$) denotes which ensembles (Base and \LGCNN{}) contain the best individual models according to validation loss.}
    \label{tab:results_1x1_uww}
    \scriptsize
    \begin{ruledtabular}
        \begin{tabular}{l | l | l | l | l}
            & $8 \cdot 8$ & $16 \cdot 16$ & $32 \cdot 32$ & $64 \cdot 64$   \\
            \hline
            Variance & \nnum{5.97e-02} & \nnum{5.69e-02} & \nnum{5.64e-02} & \nnum{5.61e-02} \\
            \hline
            \LGCNN{}$^*$& \bnum{2.19e-08} & \nnum{2.19e-08} & \nnum{2.19e-08} & \nnum{2.19e-08}  \\ 
            \hline 
            Base S1 (tanh)& \nnum{8.82e-06} & \nnum{3.43e-06} & \nnum{1.98e-06} & \nnum{1.53e-06}  \\ 
            Base S1 (sigm)& \nnum{5.07e-06} & \nnum{1.60e-06} & \nnum{7.89e-07} & \nnum{5.93e-07}  \\ 
            Base S1 (leaky)& \nnum{2.26e-08} & \nnum{8.23e-09} & \nnum{4.55e-09} & \nnum{3.42e-09}  \\ 
            Base S1 (relu)& \nnum{3.14e-08} & \bnum{6.99e-09} & \bnum{2.55e-09} & \bnum{1.20e-09}  \\ 
            \hline 
            Base S2 (tanh)& \nnum{1.57e-05} & \nnum{4.87e-06} & \nnum{2.19e-06} & \nnum{1.37e-06}  \\ 
            Base S2 (sigm)& \nnum{4.74e-06} & \nnum{1.84e-06} & \nnum{9.49e-07} & \nnum{7.83e-07}  \\ 
            Base S2 (leaky)$^*$& \nnum{2.67e-07} & \nnum{3.40e-08} & \nnum{1.47e-08} & \nnum{9.15e-09}  \\ 
            Base S2 (relu)& \nnum{4.82e-07} & \nnum{2.21e-07} & \nnum{6.37e-08} & \nnum{4.23e-08}  \\ 
            \hline 
            Base S3 (tanh)& \nnum{1.52e-06} & \nnum{5.78e-07} & \nnum{3.21e-07} & \nnum{2.48e-07}  \\ 
            Base S3 (sigm)& \nnum{1.40e-06} & \nnum{3.86e-07} & \nnum{1.27e-07} & \nnum{7.98e-08}  \\ 
            Base S3 (leaky)& \nnum{1.10e-07} & \nnum{3.78e-08} & \nnum{2.47e-08} & \nnum{1.83e-08}  \\ 
            Base S3 (relu)& \nnum{5.94e-07} & \nnum{2.25e-07} & \nnum{6.05e-08} & \nnum{4.95e-08}  \\ 
            \hline 
            Base M1 (tanh)& \nnum{1.39e-05} & \nnum{4.35e-06} & \nnum{1.62e-06} & \nnum{9.62e-07}  \\ 
            Base M1 (sigm)& \nnum{9.52e-06} & \nnum{3.76e-06} & \nnum{2.54e-06} & \nnum{2.28e-06}  \\ 
            Base M1 (leaky)& \nnum{2.41e-07} & \nnum{8.54e-08} & \nnum{4.93e-08} & \nnum{2.61e-08}  \\ 
            Base M1 (relu)& \nnum{5.22e-06} & \nnum{1.35e-06} & \nnum{6.63e-07} & \nnum{4.19e-07}  \\ 
            \hline 
            Base M2 (tanh)& \nnum{1.33e-05} & \nnum{4.27e-06} & \nnum{1.79e-06} & \nnum{1.08e-06}  \\ 
            Base M2 (sigm)& \nnum{3.86e-06} & \nnum{1.38e-06} & \nnum{7.22e-07} & \nnum{6.32e-07}  \\ 
            Base M2 (leaky)& \nnum{7.21e-08} & \nnum{1.83e-08} & \nnum{8.10e-09} & \nnum{4.81e-09}  \\ 
            Base M2 (relu)& \nnum{2.24e-06} & \nnum{7.41e-07} & \nnum{2.03e-07} & \nnum{1.25e-07}  \\ 
            \hline 
            Base M3 (tanh)& \nnum{7.21e-06} & \nnum{3.31e-06} & \nnum{2.46e-06} & \nnum{1.69e-06}  \\ 
            Base M3 (sigm)& \nnum{2.51e-06} & \nnum{9.12e-07} & \nnum{5.62e-07} & \nnum{5.02e-07}  \\ 
            Base M3 (leaky)& \nnum{1.17e-07} & \nnum{3.39e-08} & \nnum{1.23e-08} & \nnum{7.25e-09}  \\ 
            Base M3 (relu)& \nnum{6.51e-07} & \nnum{2.47e-07} & \nnum{4.64e-08} & \nnum{1.57e-08}  \\ 
            \hline 
            Base L1 (tanh)& \nnum{1.12e-05} & \nnum{3.71e-06} & \nnum{1.73e-06} & \nnum{1.04e-06}  \\ 
            Base L1 (sigm)& \nnum{5.97e-02} & \nnum{5.69e-02} & \nnum{5.64e-02} & \nnum{5.61e-02}  \\ 
            Base L1 (leaky)& \nnum{7.83e-07} & \nnum{1.89e-07} & \nnum{7.43e-08} & \nnum{5.22e-08}  \\ 
            Base L1 (relu)& \nnum{4.12e-06} & \nnum{9.68e-07} & \nnum{2.97e-07} & \nnum{1.37e-07}  \\ 
            \hline 
            Base L2 (tanh)& \nnum{1.91e-05} & \nnum{6.74e-06} & \nnum{2.88e-06} & \nnum{2.11e-06}  \\ 
            Base L2 (sigm)& \nnum{5.97e-02} & \nnum{5.69e-02} & \nnum{5.64e-02} & \nnum{5.61e-02}  \\ 
            Base L2 (leaky)& \nnum{3.00e-06} & \nnum{1.31e-06} & \nnum{9.96e-07} & \nnum{9.52e-07}  \\ 
            Base L2 (relu)& \nnum{3.21e-06} & \nnum{8.96e-07} & \nnum{2.86e-07} & \nnum{1.83e-07}  \\ 
            \hline 
            Base L3 (tanh)& \nnum{1.07e-05} & \nnum{3.38e-06} & \nnum{1.69e-06} & \nnum{1.35e-06}  \\ 
            Base L3 (sigm)& \nnum{2.34e-05} & \nnum{7.51e-06} & \nnum{2.90e-06} & \nnum{2.37e-06}  \\ 
            Base L3 (leaky)& \nnum{2.90e-06} & \nnum{7.75e-07} & \nnum{3.34e-07} & \nnum{1.73e-07}  \\ 
            Base L3 (relu)& \nnum{2.66e-06} & \nnum{8.01e-07} & \nnum{2.96e-07} & \nnum{1.41e-07}  \\ 
            \hline 
            Base W1 (tanh)& \nnum{7.26e-07} & \nnum{2.31e-07} & \nnum{7.88e-08} & \nnum{4.09e-08}  \\ 
            Base W1 (sigm)& \nnum{2.23e-06} & \nnum{6.94e-07} & \nnum{2.54e-07} & \nnum{1.58e-07}  \\ 
            Base W1 (leaky)& \nnum{8.99e-07} & \nnum{2.52e-07} & \nnum{8.42e-08} & \nnum{4.33e-08}  \\ 
            Base W1 (relu)& \nnum{1.68e-06} & \nnum{4.39e-07} & \nnum{1.50e-07} & \nnum{5.44e-08}  \\ 
            \hline 
            Base W2 (tanh)& \nnum{2.50e-04} & \nnum{6.50e-05} & \nnum{2.05e-05} & \nnum{1.05e-05}  \\ 
            Base W2 (sigm)& \nnum{5.97e-02} & \nnum{5.69e-02} & \nnum{5.64e-02} & \nnum{5.61e-02}  \\ 
            Base W2 (leaky)& \nnum{6.97e-05} & \nnum{1.77e-05} & \nnum{4.83e-06} & \nnum{1.59e-06}  \\ 
            Base W2 (relu)& \nnum{5.97e-02} & \nnum{5.69e-02} & \nnum{5.64e-02} & \nnum{5.61e-02}  \\ 
            \hline 
            Base W3 (tanh)& \nnum{2.70e-05} & \nnum{7.14e-06} & \nnum{3.41e-06} & \nnum{2.10e-06}  \\ 
            Base W3 (sigm)& \nnum{5.97e-02} & \nnum{5.69e-02} & \nnum{5.64e-02} & \nnum{5.62e-02}  \\ 
            Base W3 (leaky)& \nnum{5.03e-05} & \nnum{1.50e-05} & \nnum{7.16e-06} & \nnum{4.95e-06}  \\ 
            Base W3 (relu)& \nnum{3.25e-06} & \nnum{9.34e-07} & \nnum{3.44e-07} & \nnum{1.95e-07}  \\ 
        \end{tabular}
    \end{ruledtabular}
\end{table}

Test results for both baseline and \LGCNN{} ensembles are shown in tables~\ref{tab:results_1x1_u} to~\ref{tab:results_4d}. As stated  previously, we train every individual architecture multiple times to generate model ensembles. This allows us to counteract the stochastic nature of random initializations and optimizers. Our tables show the median mean squared error (MSE) for every model ensemble, evaluated on various lattice sizes. We prefer the median MSE over the mean MSE as it is less affected by statistical outliers (either low or high). The lowest values are highlighted in boldface to indicate the best performing architectures. In addition to the values of the test MSE, we also show the label variance of the test data sets. Here, the label variance refers to the statistical variance of the lattice averaged observables used in our regression tasks. Dividing the MSE by the variance yields the normalized MSE, which may allow for better comparison across different regression tasks and lattice sizes. More specifically, if a network can not extract any meaningful information during training, a local optimum can still be found by simply predicting the mean of the training set. In that case, the MSE reduces to the variance of the training set. If the statistical distributions of training and test sets are similar, the test MSE then also approximates the test set variance  and the normalized MSE reaches values close to one. However, care needs to be taken when comparing MSE and variance for datasets with inherently small variance, which is the case for some of our datasets with larger lattice sizes. Even though the traced Wilson loops exhibit strong fluctuations from lattice site to lattice site, computing the lattice average can effectively reduce these fluctuations and lead to decreasing label variance with growing lattice size. This can be observed in our 1+1D $W^{(4 \times 4)}$ dataset.

Tables~\ref{tab:results_1x1_u}, \ref{tab:results_1x1_uw} and~\ref{tab:results_1x1_uww} show the performance of baseline models on the $W^{(1 \times 1)}$ for different types of input. It is evident that baseline models strongly benefit from providing plaquettes $\mathcal{W}$ (single or both orientations) in addition to links $\mathcal{U}$ in the input layer. It is also interesting that for larger lattice sizes, the trivial \LGCNN{} architecture described in table~\ref{tab:arch_lcnn_2d} is outperformed by the baseline models, although the median MSE only differs by an order of magnitude at most. Based on these results, it is justified to always provide the baseline models with the most information possible ($\mathcal{U}$ and both orientations of $\mathcal{W}$).

Table~\ref{tab:results_mmse_w1x2_2d} shows the outcome of our comparison study for the $W^{(1 \times 2)}$ Wilson loop. In this first non-trivial example, meaning that the target label is not provided in the input layer as in the case of \mbox{$1 \times 1$} loops, the benefits of using \LGCNN{}s instead of traditional CNNs become apparent: the smallest \LGCNN{} architecture outperforms every other baseline architecture by four or more orders of magnitude. Interestingly, the baseline models seem to benefit from larger lattice sizes. It is possible that this reduction in MSE can be traced back to the use of the global average in the final layers of the baseline architectures. As the lattice size grows, fluctuations in the baseline predictions seem to be reduced, leading to an overall better agreement with test data. On the other hand, this effect seems entirely absent in the case of \LGCNN{} models.

Tables~\ref{tab:results_mmse_w2x2_2d} and~\ref{tab:results_mmse_w4x4_2d} show similar results for $W^{(2 \times 2)}$ and $W^{(4 \times 4)}$ Wilson loops. For these larger loops, slightly larger \LGCNN{} models seem to perform best compared to the smallest and largest architectures. As before, \LGCNN{}s outperform traditional CNNs at all lattice sizes. 

\begin{table}[htbp]
    \caption{Test results for \LGCNN{} and baseline CNN architectures (denoted as ``Base'') on the $W^{(1\times 2)}$ regression task in 1+1D. This table shows the median MSE of the model ensemble for each architecture and each lattice size. Lowest values are highlighted in boldface. For baseline CNNs, we also add the type of activation function used. Architecture details are provided in tables~\ref{tab:arch_base_w1x2} and~\ref{tab:arch_lcnn_2d}. The asterisk ($*$) denotes which ensembles (Base and \LGCNN{}) contain the best individual models according to validation loss.}
    \label{tab:results_mmse_w1x2_2d}
    \scriptsize
    \begin{ruledtabular}
        \begin{tabular}{l | l | l | l | l}
             & $8 \cdot 8$ & $16 \cdot 16$ & $32 \cdot 32$  & $64 \cdot 64$   \\
            \hline 
            Variance & \nnum{4.50e-02} & \nnum{4.16e-02} & \nnum{4.08e-02} & \nnum{4.07e-02} \\
            \hline
            \LGCNN{} S& \bnum{7.58e-09} & \bnum{7.15e-09} & \bnum{6.99e-09} & \bnum{6.97e-09}  \\ 
            \LGCNN{} M& \nnum{1.15e-08} & \nnum{1.10e-08} & \nnum{1.08e-08} & \nnum{1.08e-08}  \\ 
            \LGCNN{} L$^*$& \nnum{1.66e-08} & \nnum{1.60e-08} & \nnum{1.57e-08} & \nnum{1.57e-08}  \\ 
            \hline 
            Base S1 (tanh)& \nnum{2.34e-03} & \nnum{6.24e-04} & \nnum{1.63e-04} & \nnum{6.52e-05}  \\ 
            Base S1 (sigm)& \nnum{2.25e-03} & \nnum{5.96e-04} & \nnum{1.62e-04} & \nnum{6.29e-05}  \\ 
            Base S1 (leaky)& \nnum{2.20e-03} & \nnum{5.59e-04} & \nnum{1.45e-04} & \nnum{4.59e-05}  \\ 
            Base S1 (relu)& \nnum{2.17e-03} & \nnum{5.59e-04} & \nnum{1.50e-04} & \nnum{5.32e-05}  \\ 
            \hline 
            Base S2 (tanh)& \nnum{2.32e-03} & \nnum{6.05e-04} & \nnum{1.65e-04} & \nnum{7.00e-05}  \\ 
            Base S2 (sigm)& \nnum{2.23e-03} & \nnum{5.85e-04} & \nnum{1.52e-04} & \nnum{5.22e-05}  \\ 
            Base S2 (leaky)& \nnum{2.14e-03} & \nnum{5.55e-04} & \nnum{1.44e-04} & \nnum{5.33e-05}  \\ 
            Base S2 (relu)& \nnum{2.09e-03} & \nnum{5.31e-04} & \nnum{1.48e-04} & \nnum{5.38e-05}  \\ 
            \hline 
            Base S3 (tanh)& \nnum{2.19e-03} & \nnum{5.57e-04} & \nnum{1.54e-04} & \nnum{5.80e-05}  \\ 
            Base S3 (sigm)$^*$& \nnum{2.07e-03} & \nnum{5.26e-04} & \nnum{1.32e-04} & \nnum{4.29e-05}  \\ 
            Base S3 (leaky)& \nnum{2.01e-03} & \nnum{5.08e-04} & \nnum{1.39e-04} & \nnum{4.73e-05}  \\ 
            Base S3 (relu)& \nnum{2.03e-03} & \nnum{5.14e-04} & \nnum{1.38e-04} & \nnum{4.81e-05}  \\ 
            \hline 
            Base M1 (tanh)& \nnum{2.51e-03} & \nnum{6.70e-04} & \nnum{1.78e-04} & \nnum{7.41e-05}  \\ 
            Base M1 (sigm)& \nnum{2.35e-03} & \nnum{6.21e-04} & \nnum{1.61e-04} & \nnum{6.22e-05}  \\ 
            Base M1 (leaky)& \nnum{2.18e-03} & \nnum{5.63e-04} & \nnum{1.46e-04} & \nnum{5.39e-05}  \\ 
            Base M1 (relu)& \nnum{2.29e-03} & \nnum{6.04e-04} & \nnum{1.77e-04} & \nnum{7.70e-05}  \\ 
            \hline 
            Base M2 (tanh)& \nnum{2.51e-03} & \nnum{6.84e-04} & \nnum{1.83e-04} & \nnum{7.66e-05}  \\ 
            Base M2 (sigm)& \nnum{2.31e-03} & \nnum{6.04e-04} & \nnum{1.53e-04} & \nnum{5.46e-05}  \\ 
            Base M2 (leaky)& \nnum{2.11e-03} & \nnum{5.33e-04} & \nnum{1.38e-04} & \nnum{4.87e-05}  \\ 
            Base M2 (relu)& \nnum{2.16e-03} & \nnum{5.52e-04} & \nnum{1.42e-04} & \nnum{5.36e-05}  \\ 
            \hline 
            Base M3 (tanh)& \nnum{2.89e-03} & \nnum{7.80e-04} & \nnum{2.24e-04} & \nnum{1.10e-04}  \\ 
            Base M3 (sigm)& \nnum{2.43e-03} & \nnum{6.39e-04} & \nnum{1.76e-04} & \nnum{6.92e-05}  \\ 
            Base M3 (leaky)& \nnum{2.31e-03} & \nnum{6.02e-04} & \nnum{1.53e-04} & \nnum{5.35e-05}  \\ 
            Base M3 (relu)& \nnum{2.43e-03} & \nnum{6.39e-04} & \nnum{1.77e-04} & \nnum{6.40e-05}  \\ 
            \hline 
            Base L1 (tanh)& \nnum{2.63e-03} & \nnum{7.16e-04} & \nnum{2.10e-04} & \nnum{9.13e-05}  \\ 
            Base L1 (sigm)& \nnum{2.38e-03} & \nnum{6.33e-04} & \nnum{1.76e-04} & \nnum{7.41e-05}  \\ 
            Base L1 (leaky)& \nnum{2.30e-03} & \nnum{5.93e-04} & \nnum{1.55e-04} & \nnum{5.49e-05}  \\ 
            Base L1 (relu)& \nnum{2.33e-03} & \nnum{6.23e-04} & \nnum{2.10e-04} & \nnum{1.14e-04}  \\ 
            \hline 
            Base L2 (tanh)& \nnum{2.83e-03} & \nnum{8.03e-04} & \nnum{2.60e-04} & \nnum{1.42e-04}  \\ 
            Base L2 (sigm)& \nnum{2.87e-03} & \nnum{7.49e-04} & \nnum{2.25e-04} & \nnum{1.12e-04}  \\ 
            Base L2 (leaky)& \nnum{2.41e-03} & \nnum{6.24e-04} & \nnum{1.72e-04} & \nnum{7.11e-05}  \\ 
            Base L2 (relu)& \nnum{2.51e-03} & \nnum{6.89e-04} & \nnum{2.32e-04} & \nnum{1.29e-04}  \\ 
            \hline 
            Base L3 (tanh)& \nnum{2.70e-03} & \nnum{7.38e-04} & \nnum{2.08e-04} & \nnum{9.14e-05}  \\ 
            Base L3 (sigm)& \nnum{2.53e-03} & \nnum{6.70e-04} & \nnum{1.87e-04} & \nnum{8.39e-05}  \\ 
            Base L3 (leaky)& \nnum{2.38e-03} & \nnum{6.55e-04} & \nnum{1.91e-04} & \nnum{8.43e-05}  \\ 
            Base L3 (relu)& \nnum{2.36e-03} & \nnum{6.38e-04} & \nnum{1.81e-04} & \nnum{7.93e-05}  \\ 
            \hline 
            Base W1 (tanh)& \nnum{3.73e-03} & \nnum{1.49e-03} & \nnum{8.38e-04} & \nnum{6.60e-04}  \\ 
            Base W1 (sigm)& \nnum{2.39e-03} & \nnum{6.35e-04} & \nnum{1.72e-04} & \nnum{6.35e-05}  \\ 
            Base W1 (leaky)& \nnum{2.14e-03} & \nnum{5.69e-04} & \nnum{1.78e-04} & \nnum{9.01e-05}  \\ 
            Base W1 (relu)& \nnum{2.14e-03} & \nnum{5.41e-04} & \nnum{1.55e-04} & \nnum{6.43e-05}  \\ 
            \hline 
            Base W2 (tanh)& \nnum{2.76e-03} & \nnum{7.59e-04} & \nnum{2.09e-04} & \nnum{9.09e-05}  \\ 
            Base W2 (sigm)& \nnum{4.50e-02} & \nnum{4.16e-02} & \nnum{4.09e-02} & \nnum{4.07e-02}  \\ 
            Base W2 (leaky)& \nnum{2.37e-03} & \nnum{6.75e-04} & \nnum{1.81e-04} & \nnum{7.37e-05}  \\ 
            Base W2 (relu)& \nnum{4.50e-02} & \nnum{4.16e-02} & \nnum{4.09e-02} & \nnum{4.07e-02}  \\ 
            \hline 
            Base W3 (tanh)& \nnum{2.81e-03} & \nnum{9.02e-04} & \nnum{4.35e-04} & \nnum{3.15e-04}  \\ 
            Base W3 (sigm)& \nnum{4.50e-02} & \nnum{4.16e-02} & \nnum{4.09e-02} & \nnum{4.07e-02}  \\ 
            Base W3 (leaky)& \nnum{2.51e-03} & \nnum{6.94e-04} & \nnum{2.41e-04} & \nnum{1.35e-04}  \\ 
            Base W3 (relu)& \nnum{2.12e-03} & \nnum{5.55e-04} & \nnum{1.78e-04} & \nnum{8.62e-05}  \\ 
        \end{tabular}
    \end{ruledtabular}
\end{table}

\begin{table}
    \caption{Test results for \LGCNN{} and baseline CNN architectures (denoted as ``Base'') on the $W^{(2\times 2)}$ regression task in 1+1D. We use the same notation as in table~\ref{tab:results_1x1_u}. Architecture details are provided in tables~\ref{tab:arch_base_w2x2} and~\ref{tab:arch_lcnn_2d}. The asterisk ($*$) denotes which ensembles (Base and \LGCNN{}) contain the best individual models according to validation loss.}
    \label{tab:results_mmse_w2x2_2d}
    \scriptsize
    \begin{ruledtabular}
        \begin{tabular}{l | l | l | l | l}
             & $8 \cdot 8$ & $16 \cdot 16$ & $32 \cdot 32$  & $64 \cdot 64$   \\
            \hline
            Variance & \nnum{1.96e-02} & \nnum{1.55e-02} & \nnum{1.47e-02} & \nnum{1.45e-02} \\
            \hline
            \LGCNN{} S& \nnum{1.17e-07} & \nnum{6.91e-08} & \nnum{6.79e-08} & \nnum{6.77e-08}  \\ 
            \LGCNN{} M$^*$& \bnum{3.24e-08} & \bnum{1.96e-08} & \bnum{1.68e-08} & \bnum{1.64e-08}  \\ 
            \LGCNN{} L& \nnum{6.67e-08} & \nnum{3.89e-08} & \nnum{3.18e-08} & \nnum{3.02e-08}  \\ 
            \hline 
            Base S1 (tanh)& \nnum{4.15e-03} & \nnum{1.10e-03} & \nnum{3.15e-04} & \nnum{1.27e-04}  \\ 
            Base S1 (sigm)& \nnum{3.88e-03} & \nnum{9.98e-04} & \nnum{2.81e-04} & \nnum{9.60e-05}  \\ 
            Base S1 (leaky)& \nnum{3.88e-03} & \nnum{1.01e-03} & \nnum{2.91e-04} & \nnum{1.11e-04}  \\ 
            Base S1 (relu)& \nnum{3.93e-03} & \nnum{1.01e-03} & \nnum{2.96e-04} & \nnum{1.09e-04}  \\ 
            \hline 
            Base S2 (tanh)& \nnum{3.80e-03} & \nnum{9.75e-04} & \nnum{2.82e-04} & \nnum{1.04e-04}  \\ 
            Base S2 (sigm)& \nnum{3.82e-03} & \nnum{1.00e-03} & \nnum{2.95e-04} & \nnum{1.12e-04}  \\ 
            Base S2 (leaky)$^*$& \nnum{3.71e-03} & \nnum{9.54e-04} & \nnum{2.61e-04} & \nnum{8.63e-05}  \\ 
            Base S2 (relu)& \nnum{3.86e-03} & \nnum{9.89e-04} & \nnum{2.77e-04} & \nnum{1.00e-04}  \\ 
            \hline 
            Base S3 (tanh)& \nnum{4.15e-03} & \nnum{1.11e-03} & \nnum{3.20e-04} & \nnum{1.26e-04}  \\ 
            Base S3 (sigm)& \nnum{3.85e-03} & \nnum{9.74e-04} & \nnum{2.60e-04} & \nnum{8.31e-05}  \\ 
            Base S3 (leaky)& \nnum{3.89e-03} & \nnum{1.02e-03} & \nnum{2.93e-04} & \nnum{1.14e-04}  \\ 
            Base S3 (relu)& \nnum{3.86e-03} & \nnum{1.01e-03} & \nnum{2.86e-04} & \nnum{1.06e-04}  \\ 
            \hline 
            Base M1 (tanh)& \nnum{4.19e-03} & \nnum{1.08e-03} & \nnum{3.08e-04} & \nnum{1.21e-04}  \\ 
            Base M1 (sigm)& \nnum{3.98e-03} & \nnum{1.04e-03} & \nnum{2.93e-04} & \nnum{1.04e-04}  \\ 
            Base M1 (leaky)& \nnum{3.87e-03} & \nnum{9.96e-04} & \nnum{2.77e-04} & \nnum{1.02e-04}  \\ 
            Base M1 (relu)& \nnum{4.13e-03} & \nnum{1.11e-03} & \nnum{3.42e-04} & \nnum{1.66e-04}  \\ 
            \hline 
            Base M2 (tanh)& \nnum{4.20e-03} & \nnum{1.14e-03} & \nnum{3.61e-04} & \nnum{1.77e-04}  \\ 
            Base M2 (sigm)& \nnum{3.99e-03} & \nnum{1.06e-03} & \nnum{2.95e-04} & \nnum{1.21e-04}  \\ 
            Base M2 (leaky)& \nnum{3.94e-03} & \nnum{1.02e-03} & \nnum{2.83e-04} & \nnum{1.01e-04}  \\ 
            Base M2 (relu)& \nnum{4.18e-03} & \nnum{1.11e-03} & \nnum{3.51e-04} & \nnum{1.68e-04}  \\ 
            \hline 
            Base M3 (tanh)& \nnum{4.57e-03} & \nnum{1.19e-03} & \nnum{3.72e-04} & \nnum{1.82e-04}  \\ 
            Base M3 (sigm)& \nnum{4.26e-03} & \nnum{1.15e-03} & \nnum{3.62e-04} & \nnum{1.71e-04}  \\ 
            Base M3 (leaky)& \nnum{4.04e-03} & \nnum{1.05e-03} & \nnum{2.95e-04} & \nnum{1.12e-04}  \\ 
            Base M3 (relu)& \nnum{4.47e-03} & \nnum{1.21e-03} & \nnum{4.29e-04} & \nnum{2.30e-04}  \\ 
            \hline 
            Base L1 (tanh)& \nnum{4.26e-03} & \nnum{1.13e-03} & \nnum{3.21e-04} & \nnum{1.26e-04}  \\ 
            Base L1 (sigm)& \nnum{1.96e-02} & \nnum{1.55e-02} & \nnum{1.47e-02} & \nnum{1.45e-02}  \\ 
            Base L1 (leaky)& \nnum{4.01e-03} & \nnum{1.07e-03} & \nnum{3.14e-04} & \nnum{1.25e-04}  \\ 
            Base L1 (relu)& \nnum{1.96e-02} & \nnum{1.55e-02} & \nnum{1.47e-02} & \nnum{1.45e-02}  \\ 
            \hline 
            Base L2 (tanh)& \nnum{4.19e-03} & \nnum{1.12e-03} & \nnum{3.43e-04} & \nnum{1.54e-04}  \\ 
            Base L2 (sigm)& \nnum{1.96e-02} & \nnum{1.55e-02} & \nnum{1.47e-02} & \nnum{1.45e-02}  \\ 
            Base L2 (leaky)& \nnum{4.07e-03} & \nnum{1.07e-03} & \nnum{3.32e-04} & \nnum{1.54e-04}  \\ 
            Base L2 (relu)& \nnum{4.26e-03} & \nnum{1.14e-03} & \nnum{3.50e-04} & \nnum{1.69e-04}  \\ 
            \hline 
            Base L3 (tanh)& \nnum{4.34e-03} & \nnum{1.17e-03} & \nnum{3.60e-04} & \nnum{1.70e-04}  \\ 
            Base L3 (sigm)& \nnum{1.96e-02} & \nnum{1.55e-02} & \nnum{1.47e-02} & \nnum{1.45e-02}  \\ 
            Base L3 (leaky)& \nnum{4.07e-03} & \nnum{1.08e-03} & \nnum{3.18e-04} & \nnum{1.38e-04}  \\ 
            Base L3 (relu)& \nnum{1.96e-02} & \nnum{1.55e-02} & \nnum{1.47e-02} & \nnum{1.45e-02}  \\ 
        \end{tabular}
    \end{ruledtabular}
\end{table}

\begin{table}[htbp]
    \caption{Test results for \LGCNN{} and baseline CNN architectures (denoted as ``Base'') on the $W^{(4\times 4)}$ regression task in 1+1D. We use the same notation as in table~\ref{tab:results_1x1_u}. Architecture details are provided in tables~\ref{tab:arch_base_w4x4} and~\ref{tab:arch_lcnn_2d}. The asterisk ($*$) denotes which ensembles (Base and \LGCNN{}) contain the best individual models according to validation loss.}
    \label{tab:results_mmse_w4x4_2d}
    \scriptsize
    \begin{ruledtabular}
        \begin{tabular}{l | l | l | l | l}
            & $8 \cdot 8$ & $16 \cdot 16$ & $32 \cdot 32$  & $64 \cdot 64$   \\
            \hline
            Variance & \nnum{4.79e-03} & \nnum{1.14e-03} & \nnum{2.97e-04} & \nnum{8.53e-05} \\
            \hline
            \LGCNN{} S$^*$& \nnum{3.34e-07} & \nnum{1.51e-07} & \nnum{1.17e-07} & \nnum{1.06e-07}  \\ 
            \LGCNN{} M& \bnum{2.06e-07} & \bnum{7.15e-08} & \bnum{4.00e-08} & \bnum{3.10e-08}  \\ 
            \LGCNN{} L& \nnum{2.82e-07} & \nnum{1.09e-07} & \nnum{6.11e-08} & \nnum{5.26e-08}  \\ 
            \hline 
            Base S1 (tanh)& \nnum{4.80e-03} & \nnum{1.15e-03} & \nnum{2.95e-04} & \nnum{8.52e-05}  \\ 
            Base S1 (sigm)& \nnum{4.79e-03} & \nnum{1.14e-03} & \nnum{2.88e-04} & \nnum{7.88e-05}  \\ 
            Base S1 (leaky)& \nnum{4.79e-03} & \nnum{1.13e-03} & \nnum{2.89e-04} & \nnum{7.88e-05}  \\ 
            Base S1 (relu)$^*$& \nnum{4.79e-03} & \nnum{1.14e-03} & \nnum{2.97e-04} & \nnum{8.53e-05}  \\ 
            \hline 
            Base S2 (tanh)& \nnum{4.80e-03} & \nnum{1.14e-03} & \nnum{2.95e-04} & \nnum{8.35e-05}  \\ 
            Base S2 (sigm)& \nnum{4.80e-03} & \nnum{1.13e-03} & \nnum{2.89e-04} & \nnum{7.97e-05}  \\ 
            Base S2 (leaky)& \nnum{4.79e-03} & \nnum{1.14e-03} & \nnum{2.92e-04} & \nnum{8.16e-05}  \\ 
            Base S2 (relu)& \nnum{4.79e-03} & \nnum{1.14e-03} & \nnum{2.92e-04} & \nnum{8.09e-05}  \\ 
            \hline 
            Base S3 (tanh)& \nnum{4.80e-03} & \nnum{1.14e-03} & \nnum{2.92e-04} & \nnum{8.13e-05}  \\ 
            Base S3 (sigm)& \nnum{4.80e-03} & \nnum{1.14e-03} & \nnum{2.91e-04} & \nnum{8.03e-05}  \\ 
            Base S3 (leaky)& \nnum{4.80e-03} & \nnum{1.14e-03} & \nnum{2.94e-04} & \nnum{8.28e-05}  \\ 
            Base S3 (relu)& \nnum{4.80e-03} & \nnum{1.14e-03} & \nnum{2.92e-04} & \nnum{8.20e-05}  \\ 
            \hline 
            Base M1 (tanh)& \nnum{4.81e-03} & \nnum{1.15e-03} & \nnum{2.93e-04} & \nnum{8.22e-05}  \\ 
            Base M1 (sigm)& \nnum{4.79e-03} & \nnum{1.14e-03} & \nnum{2.95e-04} & \nnum{8.30e-05}  \\ 
            Base M1 (leaky)& \nnum{4.80e-03} & \nnum{1.14e-03} & \nnum{2.90e-04} & \nnum{8.02e-05}  \\ 
            Base M1 (relu)& \nnum{4.79e-03} & \nnum{1.14e-03} & \nnum{2.94e-04} & \nnum{8.40e-05}  \\ 
            \hline 
            Base M2 (tanh)& \nnum{4.80e-03} & \nnum{1.14e-03} & \nnum{2.93e-04} & \nnum{8.15e-05}  \\ 
            Base M2 (sigm)& \nnum{4.79e-03} & \nnum{1.14e-03} & \nnum{2.96e-04} & \nnum{8.33e-05}  \\ 
            Base M2 (leaky)& \nnum{4.80e-03} & \nnum{1.14e-03} & \nnum{2.99e-04} & \nnum{8.76e-05}  \\ 
            Base M2 (relu)& \nnum{4.79e-03} & \nnum{1.14e-03} & \nnum{2.97e-04} & \nnum{8.54e-05}  \\ 
            \hline 
            Base M3 (tanh)& \nnum{4.79e-03} & \nnum{1.14e-03} & \nnum{2.94e-04} & \nnum{8.45e-05}  \\ 
            Base M3 (sigm)& \nnum{4.79e-03} & \nnum{1.14e-03} & \nnum{2.97e-04} & \nnum{8.55e-05}  \\ 
            Base M3 (leaky)& \nnum{4.79e-03} & \nnum{1.14e-03} & \nnum{2.99e-04} & \nnum{8.68e-05}  \\ 
            Base M3 (relu)& \nnum{4.79e-03} & \nnum{1.14e-03} & \nnum{2.97e-04} & \nnum{8.53e-05}  \\ 
            \hline 
            Base L1 (tanh)& \nnum{4.83e-03} & \nnum{1.14e-03} & \nnum{2.99e-04} & \nnum{8.74e-05}  \\ 
            Base L1 (sigm)& \nnum{4.79e-03} & \nnum{1.14e-03} & \nnum{2.97e-04} & \nnum{8.55e-05}  \\ 
            Base L1 (leaky)& \nnum{4.79e-03} & \nnum{1.13e-03} & \nnum{2.92e-04} & \nnum{8.16e-05}  \\ 
            Base L1 (relu)& \nnum{4.79e-03} & \nnum{1.14e-03} & \nnum{2.97e-04} & \nnum{8.54e-05}  \\ 
            \hline 
            Base L2 (tanh)& \nnum{4.80e-03} & \nnum{1.14e-03} & \nnum{2.95e-04} & \nnum{8.41e-05}  \\ 
            Base L2 (sigm)& \nnum{4.79e-03} & \nnum{1.14e-03} & \nnum{2.97e-04} & \nnum{8.54e-05}  \\ 
            Base L2 (leaky)& \nnum{4.79e-03} & \nnum{1.14e-03} & \nnum{2.97e-04} & \nnum{8.54e-05}  \\ 
            Base L2 (relu)& \nnum{4.79e-03} & \nnum{1.14e-03} & \nnum{2.97e-04} & \nnum{8.55e-05}  \\ 
            \hline 
            Base L3 (tanh)& \nnum{4.81e-03} & \nnum{1.15e-03} & \nnum{3.00e-04} & \nnum{8.86e-05}  \\ 
            Base L3 (sigm)& \nnum{4.79e-03} & \nnum{1.14e-03} & \nnum{2.97e-04} & \nnum{8.55e-05}  \\ 
            Base L3 (leaky)& \nnum{4.79e-03} & \nnum{1.14e-03} & \nnum{2.89e-04} & \nnum{8.20e-05}  \\ 
            Base L3 (relu)& \nnum{4.79e-03} & \nnum{1.14e-03} & \nnum{2.97e-04} & \nnum{8.54e-05}  \\ 
        \end{tabular}
    \end{ruledtabular}
\end{table}

\begin{figure}
    \centering
    \includegraphics[width=0.45\textwidth]{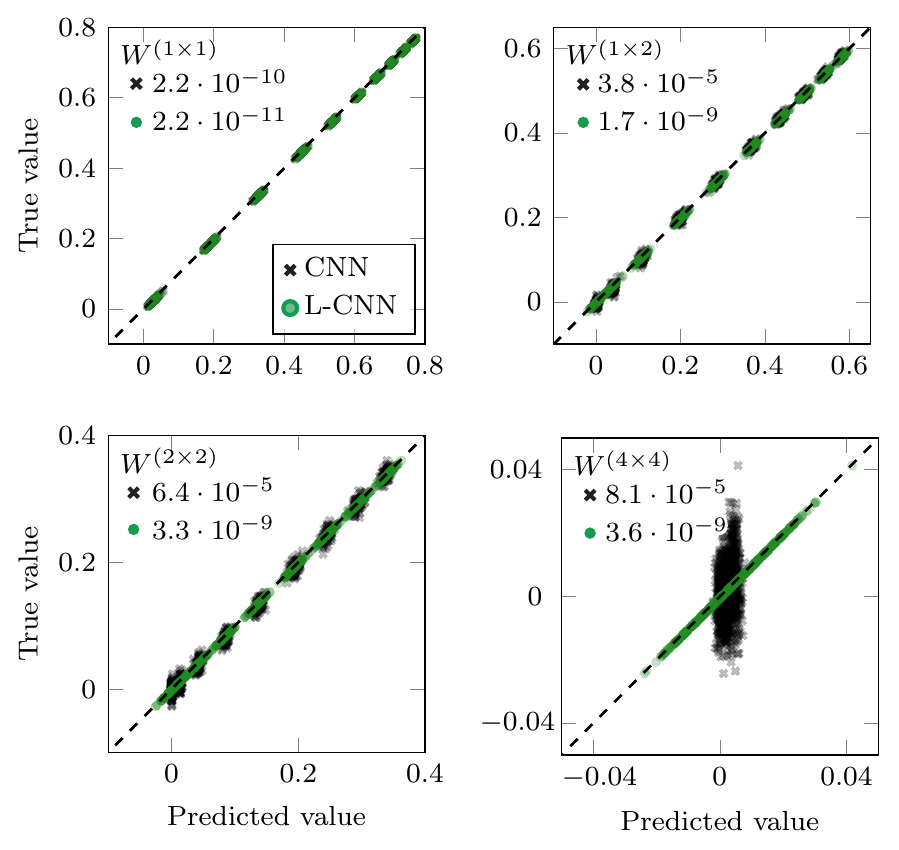}
    \caption{Scatter plots comparing best L-CNN models to baseline CNN models for Wilson loops of various sizes for 1+1D analogous to Fig.~3 of the accompanying Letter, but for the \mbox{$N_s \cdot N_t = 64 \cdot 64$} test dataset. For each configuration in the dataset, we plot the prediction and the true value (label). Perfect agreement
    is indicated by the dashed $45^\circ$ line. The values in the upper left corner denote the MSEs of each plot. The distinct clusters shown in these plots correspond to different values of the coupling constant $\beta$ used in the datasets.}
    \label{fig:d2_scatter_large}
\end{figure}

\begin{table}[htbp]
    \caption{Test results for selected $W^{(1 \times 2)}$ and $W^{(2 \times 2)}$ baseline architectures with or without global average pooling (GAP). Test MSEs for baselines with GAP are taken from tables~\ref{tab:results_mmse_w1x2_2d} and~\ref{tab:results_mmse_w2x2_2d}. In all studied cases, the median MSE is lower for baseline architectures with GAP.}
    \label{tab:results_nogap}
    \scriptsize
    \begin{ruledtabular}
        \begin{tabular}{l | l | l | l | l}
             & $8 \cdot 8$ & $16 \cdot 16$ & $32 \cdot 32$  & $64 \cdot 64$   \\
            \hline
            $W^{(1 \times 2)}$ & & & & \\
            \hline
            Variance & \nnum{4.50e-02} & \nnum{4.16e-02} & \nnum{4.08e-02} & \nnum{4.07e-02} \\
            \hline
            S3 (leaky) (no GAP)& \nnum{3.82e-03} & \nnum{2.07e-03} & \nnum{1.63e-03} & \nnum{1.52e-03}  \\
            S3 (leaky) (GAP) & \bnum{2.01e-03} & \bnum{5.08e-04} & \bnum{1.39e-04} & \bnum{4.73e-05}  \\ 
            \hline
            L2 (tanh) (no GAP)& \nnum{3.44e-03} & \nnum{1.42e-03} & \nnum{8.96e-04} & \nnum{8.27e-04}  \\ 
            L2 (tanh) (GAP) & \bnum{2.83e-03} & \bnum{8.03e-04} & \bnum{2.60e-04} & \bnum{1.42e-04}  \\ 
            \hline
            \hline
            $W^{(2 \times 2)}$ & & & & \\
            \hline
            Variance & \nnum{1.96e-02} & \nnum{1.55e-02} & \nnum{1.47e-02} & \nnum{1.45e-02} \\
            \hline
            S2 (leaky) (no GAP)& \nnum{5.15e-03} & \nnum{2.04e-03} & \nnum{1.36e-03} & \nnum{1.21e-03}  \\ 
            S2 (leaky) (GAP) & \bnum{3.71e-03} & \bnum{9.54e-04} & \bnum{2.61e-04} & \bnum{8.63e-05}  \\ 
            \hline
            L2 (tanh) (no GAP)& \nnum{4.92e-03} & \nnum{1.46e-03} & \nnum{7.19e-04} & \nnum{5.06e-04}  \\ 
            L2 (tanh) (GAP) & \bnum{4.19e-03} & \bnum{1.12e-03} & \bnum{3.43e-04} & \bnum{1.54e-04}  \\ 
        \end{tabular}
    \end{ruledtabular}
\end{table}

\begin{table}[htbp]
    \caption{Test results for \LGCNN{} architectures on all regression tasks in 3+1D. We use the same notation as in table~\ref{tab:results_1x1_u}. Architecture details are provided in table~\ref{tab:arch_lcnn_4d}.}
    \label{tab:results_4d}
    \scriptsize
    \begin{ruledtabular}
        \begin{tabular}{l | l | l | l | l}
             & $4 \cdot 8^3$ & $6 \cdot 8^3$ & $6 \cdot 12^3$  & $8 \cdot 16^3$   \\
            \hline
            $W^{(2 \times 2)}$ & & & & \\
            Variance & \nnum{7.03e-02} & \nnum{7.08e-02} & \nnum{7.05e-02} & \nnum{7.05e-02} \\
            \LGCNN{} S& \bnum{1.64e-07} & \bnum{1.63e-07} & \bnum{1.63e-07} & \nnum{1.63e-07}  \\ 
            \LGCNN{} M& \nnum{9.16e-07} & \nnum{6.18e-07} & \nnum{2.17e-07} & \bnum{1.30e-07}  \\ 
            \hline
            $W^{(4 \times 4)}$ & & & & \\
            Variance & \nnum{2.00e-02} & \nnum{2.08e-02} & \nnum{2.04e-02} & \nnum{2.03e-02} \\
            \LGCNN{} S& \bnum{3.77e-07} & \bnum{3.79e-07} & \bnum{3.74e-07} & \bnum{3.74e-07}  \\ 
            \LGCNN{} M& \nnum{8.26e-07} & \nnum{8.16e-07} & \nnum{7.99e-07} & \nnum{7.99e-07}  \\ 
            \hline
            $Q_P$ & & & & \\
            Variance & \nnum{2.91e-07} & \nnum{1.91e-07} & \nnum{6.27e-08} & \nnum{1.87e-08} \\
            \LGCNN{} S& \nnum{3.18e-09} & \nnum{3.17e-09} & \nnum{3.17e-09} & \nnum{3.17e-09}  \\ 
        \end{tabular}
    \end{ruledtabular}
\end{table}

\begin{table}[htbp]
    \caption{Test results for L-CNN architectures on all regression tasks in 3+1D without using lattice averages. We use the same notation as in table~\ref{tab:results_1x1_u}. Architecture details are provided in table~\ref{tab:arch_lcnn_4d}.}
    \label{tab:results_4d_nogap}
    \scriptsize
    \begin{ruledtabular}
        \begin{tabular}{l | l | l | l | l}
             & $4 \cdot 8^3$ & $6 \cdot 8^3$ & $6 \cdot 12^3$  & $8 \cdot 16^3$   \\
            \hline
            $W^{(2 \times 2)}$ & & & & \\
            Variance & \nnum{2.35e-01} & \nnum{2.35e-01} & \nnum{2.35e-01} & \nnum{2.35e-01} \\
            L-CNN S& \bnum{2.10e-06} & \bnum{2.12e-06} & \bnum{2.12e-06} & \bnum{2.12e-06}  \\ 
            L-CNN M& \nnum{1.58e-03} & \nnum{1.58e-03} & \nnum{1.58e-03} & \nnum{1.58e-03}  \\ 
            \hline
            $W^{(4 \times 4)}$ & & & & \\
            Variance & \nnum{2.45e-01} & \nnum{2.45e-01} & \nnum{2.45e-01} & \nnum{2.45e-01} \\
            L-CNN S& \bnum{1.38e-05} & \bnum{1.38e-05} & \bnum{1.38e-05} & \bnum{1.38e-05}  \\ 
            L-CNN M& \nnum{1.91e-05} & \nnum{1.91e-05} & \nnum{1.91e-05} & \nnum{1.91e-05}  \\ 
            \hline
            $Q_P$ & & & & \\
            Variance & \nnum{6.59e-04} & \nnum{6.60e-04} & \nnum{6.60e-04} & \nnum{6.60e-04} \\
            L-CNN S& \nnum{3.35e-09} & \nnum{3.34e-09} & \nnum{3.34e-09} & \nnum{3.34e-09}  \\ 
    \end{tabular}
    \end{ruledtabular}
\end{table}

We have visualized the predictions of our networks using scatter plots in Fig.~3 of the Letter, where we plot predicted values against the true values of an observable. In these plots we have chosen the best individual models based on validation loss on $8 \cdot 8$ lattices (see table~\ref{tab:datasets}). The ensembles containing these best models are highlighted with an asterisk ($*$) in the result tables~\ref{tab:results_1x1_uww}, \ref{tab:results_mmse_w1x2_2d}, \ref{tab:results_mmse_w2x2_2d} and~\ref{tab:results_mmse_w4x4_2d}. Using these same models, we have also visualized the predictions on $64 \cdot 64$ lattices in Fig.~\ref{fig:d2_scatter_large}.

Table~\ref{tab:results_nogap} shows the influence of global average pooling (GAP) on test performance of baseline models. For this computational experiment we have selected two architectures (small and large) each for the $W^{(1\times 2)}$ and $W^{(2\times 2)}$ regression tasks. We observe that including GAP for baseline models leads to much better median MSE on test data, which we use as justification to include GAP in every baseline architecture of our comparison study.

Finally, table~\ref{tab:results_4d} summarizes our results for \LGCNN{} models in 3+1D. Although no comparisons to baseline networks were made, we observe similar behavior and similar median MSEs as in the two-dimensional tasks. As in the previous tables, we also report the label variance of the test datasets. We find that the $Q_P$ dataset exhibits a very small variance, which is expected for lattice averages of the topological charge density on uncooled configurations. The median test MSE of our L-CNN models for $Q_P$ is just one order of magnitude below the variance, however, as mentioned previously, comparing test MSE and variance for datasets with inherently small variance can be misleading. We therefore repeat our analysis without using lattice averages, which is possible for our L-CNN models, as they are trained without a final lattice average layer. The results are shown in table~\ref{tab:results_4d_nogap}. Removing the lattice average reveals that the median MSE is in fact many orders smaller than the label variance. Surprisingly, we also find that the L-CNN S architecture for $W^{(2 \times 2)}$ is in fact much better than the corresponding L-CNN M architecture. Comparing the best models of each ensemble still yields low test MSEs: \mbox{\nnum{2.27e-05}} for L-CNN M and \mbox{\nnum{1.23e-06}} for L-CNN S. We find that in the L-CNN M ensemble for $W^{(2 \times 2)}$ two of five models are able to perform accurate predictions for each lattice site, while the other three are only able to produce acceptable lattice averaged predictions. This result is unexpected, but also likely due to having only five models in the model ensemble.

\subsection{Investigating broken gauge symmetry}
As baseline models are a priori not gauge equivariant, we can test the robustness of their predictions against gauge transformations given by
\begin{align}
    U_{\mathbf x, \mu} \rightarrow \Omega_\mathbf{x} U_{\mathbf x, \mu} \Omega_\mathbf{x+\mu}^\dg. \label{eq:gauge_transform}
\end{align}
The labels (or true values) are unaffected by these transformations because we only consider gauge invariant observables. As a first test, we investigate the sensitivity of networks due to random gauge transformations similar to~\cite{Boyda:2020nfh}. Random SU($N_c$) matrices $\Omega_{\mathbf x}$ are obtained by generating random standard normal distributed vector components $\chi^a_{\mathbf{x}}$ for $a \in \{ 1, 2, \dots,  N_c^2 - 1\}$ at every lattice site $\mathbf x$, which are plugged into the matrix exponential:
\begin{align}
     \Omega_\mathbf{x} = \exp\left( i t^a \alpha \, \chi^a_\mathbf{x} \right), \quad \alpha > 0. \label{eq:random_gauge}
\end{align}
In the above expression $\alpha > 0$ is a parameter that controls the ``largeness'' of the random gauge transformation. We generate for every physically independent lattice configuration $\mathcal{U}$ obtained from our MCMC simulation, a set of physically equivalent, gauge transformed configurations $\{ \mathcal{U}_k \}$, $k \in \{ 1, 2, \dots, N_\mathrm{gauge}\}$, and record the predictions for each of these configurations. We perform the same computational experiment for \LGCNN{} models, which are by construction gauge invariant. Consequently, the predictions are unaffected up to numerical precision.

A second, more stringent test of gauge equivariance is possible through adversarial attacks (see e.g.~\cite{Xu:2019abc} for a review). The main idea is to find specific gauge transformations $\Omega_{\mathbf x}$ which lead to the largest possible deviation between the untransformed predictions $y_\mathrm{pred}$ and transformed predictions $y_\mathrm{trans}$, i.e.~$ |y_\mathrm{pred} - y_\mathrm{trans}| \rightarrow \mathrm{max}$, of a particular baseline network. Our strategy to find these particular gauge transformations is to use an iterative optimization method: first, we randomly initialize a gauge transformation $\Omega_\mathbf{x}$ via
\begin{align}
    \Omega_\mathbf{x} = \exp{\big( i t^a \rho^a_\mathbf{x} \big)},   
\end{align}
where $\rho^a_\mathbf{x}$ are the parameters of the gauge transformation as in Eq.~\eqref{eq:random_gauge}. Then, given a particular gauge link configuration $\mathcal{U}$ and an already trained, symmetry-breaking model (such as one of our baseline CNNs), described by the model function $h:\, \mathcal{U} \rightarrow y \in \mathbb R$, we apply our initial guess $\Omega_{\mathbf x}$ (as given by $\rho^a_\mathbf{x}$) to the lattice configuration $\mathcal{U}$ via Eq.~\eqref{eq:gauge_transform}. This yields $\mathcal{U}_\rho$ and the transformed prediction
\begin{align}
    y_\mathrm{trans}(\rho) = h(\mathcal{U}_\rho).
\end{align}
Then, in order to estimate the error bounds
\begin{align}
    \min_{\rho} \, y_\mathrm{trans} \leq  y_\mathrm{pred} \leq \max_{\rho} \, y_\mathrm{trans}
\end{align}
due to gauge symmetry breaking, we alter the transformed predictions as much as possible in two independent directions, i.e.~we optimize $y_\mathrm{trans} \rightarrow \mathrm{max}$ and $y_\mathrm{trans} \rightarrow \mathrm{min}$. For the optimization procedure, we choose the loss function
\begin{align}
    \mathcal{L}(\rho) = \pm y_\mathrm{trans}(\rho),
\end{align}
which can be viewed as a function of the gauge transformation parameters $\rho^a_\mathbf{x}$. Minimizing $\mathcal{L}(\rho)$ leads to two extrema $y_\mathrm{min}$ and $y_\mathrm{max}$ depending on the chosen sign in the loss function. We use \textit{PyTorch} to compute the gradient of $\mathcal{L}(\rho)$ with respect to the parameters $\rho^a_\mathbf{x}$. The gradient can then be used by any suitable optimizer such as \textit{AdamW} to minimize the loss function and obtain the optimal gauge transformations that lead to maximal violations of gauge invariance. The results depend on the chosen model, the initial lattice configuration $\mathcal{U}$ and also on the initial values for $\rho^a_\mathrm{x}$. Note that other loss functions could be chosen as well: if we are only interested in a large deviation from the original prediction $y_{\mathrm{pred}}$, we can maximize
\begin{align}
    \mathcal{L}(\rho) = (y_\mathrm{pred} - y_\mathrm{trans}(\rho))^2,
\end{align}
and if we want to find a gauge transformation that leads to a particular target prediction $y_\mathrm{target}$, the loss function
\begin{align}
    \mathcal{L}(\rho) = (y_\mathrm{target} - y_\mathrm{trans}(\rho))^2
\end{align}
is a suitable one to minimize.

Figure~4 of the Letter shows the results for both random gauge transformations and multiple adversarial attacks applied to our best baseline CNN and \LGCNN{} for the $W^{(1 \times 2)}$ regression task in 1+1D. For each lattice configuration we have used $200$ random gauge transformations and $5$ randomly initialized adversarial attacks each for the lower and upper bounds. We observe that random gauge transformations only lead to violations of gauge invariance of up to $16$ per cent, whereas adversarial attacks can lead to deviations of up to $79$ per cent. By construction, the predictions of \LGCNN{}s are unaffected by either method (up to relative errors of the order $10^{-6}$ due to the use of single precision on the GPU). The results of adversarial attacks indicate that random gauge transformations can underestimate the sensitivity of networks without gauge equivariance (e.g.~as used in~\cite{Boyda:2020nfh}).

\subsection{Applications to Wilson flow}
In the regression task for the topological charge $Q_P$, which is given by the lattice sum over the charge density
\begin{align}
    Q_P = \sum_{\mathbf x} q_\mathbf{x}^{\mathrm{plaq}}, \label{eq:top_charge}
\end{align}
we can also perform additional tests that go beyond merely calculating MSEs. For these models, training was performed on $4 \cdot 8^3$ lattice configurations that were obtained directly from our MCMC simulations. As an observable, $Q_P$ is usually studied at larger lattices, and even more importantly, under gauge cooling or Wilson (gradient) flow~\cite{Luscher:2010iy, Alexandrou:2017hqw}. Restricted by topology, the continuum limit of Eq.~\eqref{eq:top_charge} is integer valued. On the lattice, in particular for configurations directly obtained from MCMC simulations, one has to perform some form of smearing or cooling first. In our simulations we have implemented a Wilson flow procedure based on the Wilson action using the following link updates:
\begin{align}
    U_{\mathbf x,\mu}(\tau + \Delta \tau) &= \exp{\big( i \Delta \tau \, \omega_{\mathbf x,\mu}(\tau) \big)} U_{\mathbf x,\mu}(\tau),
\end{align}
where the algebra element in the matrix exponential is given by
\begin{align}
    \omega_{\mathbf x, \mu}(\tau) = - \sum_{|\nu|} \left[ U_{\mathbf x,\mu\nu}(\tau) \right]_{\mathrm{ah}}.
\end{align}
Here, $\tau$ is the auxiliary Wilson flow time, $\Delta \tau$ is the Wilson flow time step, $\left[ X \right]_\mathrm{ah}$ denotes the anti-Hermitian, traceless part of a matrix $X$ given by 
\begin{align}
    \left[ X \right]_\mathrm{ah} = \frac{1}{2 i} \left(X \! - \! X^\dagger \right) - \frac{1}{2 i N_c} \one \, \mathrm{Tr} \left( X \! - \! X^\dagger\right),
\end{align}
and the sum $\sum_{|\nu|}$ runs over both negative and positive directions $\nu$. As an initial condition for gradient flow, we choose a lattice configuration generated by our MCMC simulation:
\begin{align}
    U_{\mathrm x,\mu}(\tau = 0) = U_{\mathrm x, \mu}.
\end{align}
Applying these updates with small time steps $\Delta \tau$ leads to a successive reduction of the Wilson action, i.e.
\begin{align}
    S_W[ \mathcal{U}(\tau + \Delta \tau) ] < 
    S_W[ \mathcal{U}(\tau)], \qquad \Delta \tau \ll 1.
\end{align}
Under gradient flow, gauge configurations~$\mathcal{U}$ admit values for $Q_P$ closer to integers, although details depend on the lattice size and the initial lattice configuration at $\tau = 0$. In Fig.~5 of the Letter, we demonstrate that \LGCNN{} models, originally trained on unflowed configurations, reproduce the actual values of $Q_P$ under gradient flow to a high degree of accuracy on $8 \cdot 24^3$ lattices. We have used a time step of $\Delta \tau = 0.005$ to obtain these results.

\newpage

\bibliography{references.bib}